\documentclass[]{interact}

\usepackage{epstopdf}
\usepackage{subfigure}

\usepackage{url}
\usepackage{multirow}

\usepackage{adjustbox}

\usepackage{natbib}
\bibpunct[, ]{(}{)}{,}{a}{}{,}
\renewcommand\bibfont{\fontsize{10}{12}\selectfont}

\theoremstyle{plain}

\theoremstyle{definition}

\theoremstyle{remark}

\begin{document}

\articletype{RESEARCH ARTICLE}

\title{Desk-AId: Humanitarian Aid Desk Assessment with Geospatial AI for Predicting Landmine Areas}


\author{
\name{
Flavio Cirillo\textsuperscript{a}\thanks{CONTACT Flavio Cirillo. Email: flavio.cirillo@neclab.eu}, 
Gurkan Solmaz\textsuperscript{a}, 
Yi-Hsuan Peng\textsuperscript{b}, 
Christian Bizer\textsuperscript{b} and
Martin Jebens\textsuperscript{c}}
\affil{\textsuperscript{a}NEC Laboratories Europe, Heidelberg, Germany; \textsuperscript{b}University of Mannheim, Mannheim, Germany; \textsuperscript{c}International Committee of the Red Cross}
}


\maketitle

\begin{abstract}
The process of clearing areas, namely demining, starts by assessing and prioritizing potential hazardous areas (i.e., \textit{desk assessment}) to go under thorough investigation of experts, who confirm the risk and proceed with the mines clearance operations. 
This paper presents Desk-AId that supports the desk assessment phase by estimating landmine risks using geospatial data and socioeconomic information. Desk-AId uses a Geospatial AI approach specialized to landmines. The approach includes mixed data sampling strategies and context-enrichment by historical conflicts and key multi-domain facilities (e.g., buildings, roads, health sites). The proposed system addresses the issue of having only ground-truth for confirmed hazardous areas by implementing a new hard-negative data sampling strategy, where negative points are sampled in the vicinity of hazardous areas.
Experiments validate Desk-Aid in two domains for landmine risk assessment: 1) country-wide, and 2) uncharted {\em study areas}). The proposed approach increases the estimation accuracies up to 92\%, for different classification models such as RandomForest (RF), Feedforward Neural Networks (FNN), and Graph Neural Networks (GNN).

\end{abstract}

\begin{keywords}
mine detection, risk assessment, machine learning, geographical artificial intelligence
\end{keywords}

\section{Introduction}
Landmines are a blight in post-conflict regions. Since landmines are cheap to produce, easy to deploy, maintenance-free, and highly durable, massive amounts were excessively deployed during recent civil conflicts.
As of October 2021, at least 60 countries and other areas remain contaminated by antipersonnel mines~\cite{report-2021}. 
Uncleared landmines claimed more than 7000 casualties in 2020 alone, and the numbers, unfortunately, have been more or less steady year-on-year for the past 20 years~\cite{report-2021}.
Furthermore, in the post-conflict period, unexploded landmines result in not only direct victims, degradation of land and contamination of natural resources, but also socio-economic underdevelopment among the affected populations~\cite{demining-sri-lanka}. For example, lands that are marked as hazardous cannot be used for agriculture, transportation, and communication infrastructure. 

\begin{figure}[h]
 \centering
  \includegraphics[width=\linewidth]{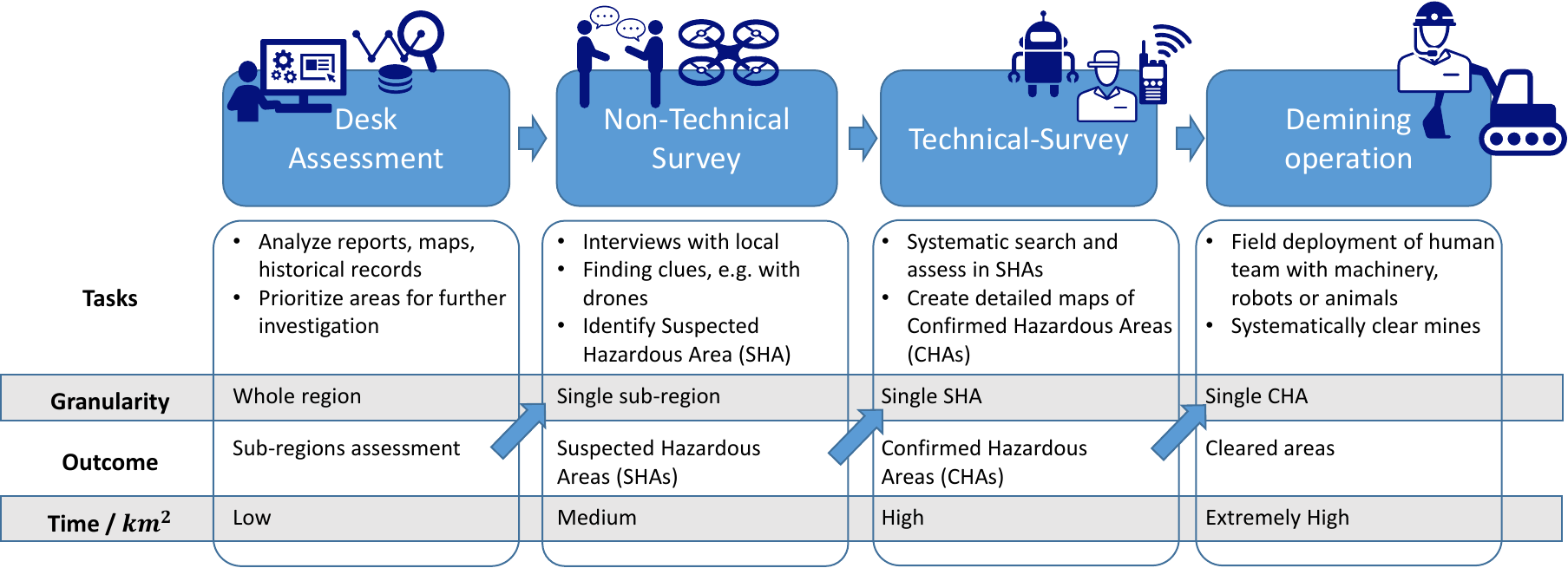}
  \caption{Break-down of demining operations into their major phases. Desk-AId falls into the phase of ``Desk Assessment'.}
  \label{fig:demining_operation}
\end{figure}

International response to the landmine problem is referred to as humanitarian mine action (HMA)~\cite{Schultz2016}.  
The purpose of the mine action is to reduce the impacts of Explosive Remnants of War (ERW)
\footnote{ERW denotes all explosive contamination from war, such as landmines and unexploded ordnance (UXO)~\cite{Lacroix2013}. In this article, the terms ERW, hazard, and landmine are used interchangeably.}
on local populations and to return cleared land to local communities for land rehabilitation.
Many global non-governmental organizations (NGOs), including International Committee of the Red Cross (ICRC), United Nations Mine Action Service (UNMAS), and the Geneva International Centre for Humanitarian Demining (GICHD), conduct demining operations that positively impact local economies and communities~\cite{schultz9}. 
Nevertheless, a significant challenge in the demining operations is the mismatch between the area's size and the available resources. 
The cost for removing a landmine can be up to 300 times more than the cost of producing it~\cite{icrc-1995}. Further, for every hour spent to deploy a mine over 1000 hours needs to be spent to demine it~\cite{burelandmine}. A team of 10 persons can manually clear in a day at maximum 500 square meters~\cite{burelandmine}.
How to effectively plan the deployment of the limited demining resources remains a persistent problem for the demining experts~\cite{Rafique2019}. 
Fig.~\ref{fig:demining_operation} shows the process of demining that starts with the desk assessment. A human expert analyzes data such as reports, historical records and maps with the aid of visualization and data tools. As results, sub-regions are assessed as potentially dangerous. Further, during this step the sub-regions are prioritized under different criteria (e.g., uncertainty of suspected areas, socio-economic importance).
The risk assessment and prioritization are essential since it can impact human life in case of false negative or the socio-economic factor (thus, human quality of life) in case of false positive.
Once potential hazardous areas are identified, a non-technical survey looks for hints to mark an area as suspected hazardous area (SHA) such as conducting interviews with local communities to gather information about the history of the area and potential mine incidents, or using technological tools such as drones~\cite{jebens2020extent}. 
A technical survey involves the deployment of expert people to the SHA to actually find proof of mine contamination using tools such as metal detectors and ground-penetrating radars. The adoption of late technology such as deep learning demonstrated also good results on speeding up this process~\cite{barnawi2022comprehensive}. 
Only after the mines have been located with precise boundaries of Confirmed Hazardous Areas (CHAs), the mines are removed and the area cleared.
Each of the described steps have a different scale in terms of focus from big to small granularity with a ratio between effort time to $km^2$ from low to very high. Therefore, it is very important to complete each step with the highest possible accuracy. 
In this article, we focus on the desk assessment, aiming at enhancing the instruments for the human domain expert to assess the landmine presence risks of sub-regions. Having an accurate risk assessments before the deployment of (technical or non-technical) people on the field is crucial to speed up the demining process of technical survey and reduce the costs.
While investigating new technologies (e.g., GIS and remote sensing) is not recent, automated landmine risk prediction systems still need to be explored. There are few studies~\cite{Rafique2019,Schultz2016} that applies ML for the risk assessment of landmine presence task.


In this work, we aim to build a pipeline for automatic landmine risk assessment in post-conflict regions. We use confirmed landmine contamination data across a whole country as starting dataset. We combine it with geographical, socio-economic, and remnants of war sectors information to build a meaningful feature set. The selected attributes have the characteristic to be easily computed from abundant data (e.g., buildings location and waterway) so as to build a system easily applicable to different regions. 
The proposed approach does not target at the precise localization of landmines but rather at the identification of potential hazard of areas. The smallest granularity considered in the proposed approach is 50 meters that is acceptable also for areas with high erosion factor after many decades as is the case of Skallingen peninsula coastline (Denmark) contaminated by mines deployed during the second world war and cleared only in 2012~\cite{jebens2013analyzing}.
Further, we address the uncertainty of negative points by sampling points in the vicinity of the CHAs that are positive points areas. The idea is that areas close to the CHAs have been more likely surveyed by technical and non-technical expert. We test different combination of distances to the CHAs for the negative sampling.
We extensively test our sampling approach in multiple scenarios such as: i) training and testing throughout whole country risk assessment, ii) testing in CHAs vicinity areas, iii) testing in completely uncharted areas.
We explore also different ML approaches such as tabular data-based and graph data-based.

Our objectives are two-fold: increase the geographic granularity for detection of hazardous area and the prediction of risk in unexplored areas. 
This work selects Afghanistan as a case study due to it being one of the countries that has suffered the most from landmines and the related ERW.

The main contributions of this paper are highlighted as follows:

\begin{itemize} 

\item \textit{Building a dataset from geographical and socio-economic data}. We develop a generic landmine detection pipeline to build mine contamination dataset across the whole country's land, as well as handling features among geographical, socio-economic, and remnants of war sectors. We also provide insight on the role played by features of different types and their relationships between each other.

\item \textit{Design an intelligent balanced data sampling method to address uncertainty issue on negative sampling.} We address the challenge of having ground-truth only for positive points by proposing a strategy to build a dataset by composing positive points and negatives points geographically close to the positive points. We leverage the idea that areas close by the ground truth positive areas have been more likely surveyed by domain experts if compared to remote locations. We test granularity of 50, 500 and 5000 meters and combinations of them.


\item \textit{Explore Graph Neural Network to exploit the nature of the geographical data.} We implement a location-based graph construction methodology for modeling the neighboring geographical location. The implemented Graph Neural Networks (GNN) outperforms the other commonly-used algorithms such as Feedforward Neural Networks (FNN)

\item \textit{Extensive experimentation of our pipeline in whole country and uncharted area scenarios.} We first test our approach assessing risk for the whole country of Afghanistan. Then, we split the dataset geographically by removing controlled areas for which we have ground-truth CHAs points from the training dataset. These controlled areas are used for testing as uncharted areas.  We select two different study areas with different characteristics: an urban area (around the capital Kabul) and an extended remote area in the center of Afghanistan. 
We also test the impact of different size of feature set. For every experiment we test various classes of ML model.

\item \textit{Visualization of the results to support decision-making.} Desk-AId is meant to augment already existing desk assessment tools. Thus, we serialize the results into importable files and show them into QGIS~\footnote{QGIS, \url{https://qgis.org}}, a free and open source geographic system, as an example visualization to support domain expert. QGIS is often adopted as visualization tools for landmine desk assessment purpose.

\end{itemize}


The final target of this work is to create a viable system that is easy to re-use and quickly applicable to a new region with different characteristics. The aim is to operationalize the desk assessment of landmine hazardous areas with minimal effort on the domain expert speeding up the demining operation in multiple countries.
\section{Related Work}
 In the following, we first introduce an emerging new research topic, namely Geospatial AI or GeoAI, that produced relevant studies to this paper. We, then, examine relevant research questions in agriculture mining. Finally we discuss current investigations that work on the landmine detection and prediction problems. 

\subsection{Geospatial AI}
Geospatial Artificial Intelligence (shortly called as ``GeoAI'') is an emerging field that leverages high-performance computing to analyze large amounts of spatial data using AI techniques such as ML, deep learning, and data mining. It combines aspects of spatial science, requiring specific technologies, such as GIS, with AI to extract meaningful insights from big data~\cite{Vopham2018}. Constant expansion of big spatial data is one of the reasons to drive GeoAI. Two prominent examples are remote sensing and volunteered geographic information (VGI), which encapsulates user-generated content with a location component. In recent years, VGI exploded with the advent and continued expansion of social media and smartphones~\cite{Hansi2017}. The OpenStreetMap (OSM)~\cite{OpenStreetMap}, that we use in this work, demonstrates the benefit of VGI: everyone can use a phone to access and annotate the map attributes.

Similar to this work, Lin et al.~\cite{Lin2017} apply the RF model and mine OSM spatial big data to select the most important geographic features (e.g., land use and roads) for their task, PM$_{2.5}$ concentration prediction. Zhu et al.~\cite{Zhu2020} demonstrates the promising use of graph convolutional neural networks (GCNNs)~\cite{Wu2019} in geographic knowledge prediction tasks. Their case study is designed as a node classification task in the Beijing metropolitan area for predicting the unobserved place characteristics (e.g., dining, residence) based on the observed properties as nodes and specific place connections as edges using GCNNs. %
They compare the result of different edges inside the graph, namely no connection, the connection between spatially adjacent places, and spatial interaction, which they incorporate a taxi traffic record between locations. Since the edge type of spatial interaction displays the best overall accuracy, they conclude that performance can be improved by using more relevant place connections and more information explanatory characteristics. Even though the geographic data at this work does not have an existing graph structure, we consider connecting adjacent places and comparing the GCNNs result with Feedforwards Neural Networks which treat each location as an independent individual. Zhang et al.~\cite{ZhangACMIMWUT} study the problem of anomaly detection of data sources in geographical regions (i.e., urban areas) using spatio-temporal data. Similar techniques can complement our work through their application in the humanitarian datasets where data source anomalies may exist.

\subsection{Agriculture Mining}
Due to the high reliance on geographic data for prediction models and enormous economic benefits, the abundant technique used in the agriculture mining domain is relevant to the ones we apply in this work.
Agriculture mining, or smart farming, is the research field that tackles the challenges of agricultural production in terms of productivity, environmental impact, food security, and sustainability~\cite{Kamilaris2017}. One of the concepts, namely precision agriculture, is the generation of spatial variability maps that employ precise localization of point measurements in the field. This is analogous in the mine action where the technical survey aims to reduce the size of the mine-contaminated area. 

Schuster et al.~\cite{Schuster2011} explore the use of a clustering algorithm ($k$-means) to identify management zones of cotton, with the dependent variable being cotton yield and the independent variable including multi-spectral imaging of the crop and physical characteristics of the field, e.g., slope, soil. The research does not, however, consider more advanced algorithms. Harshath et al.~\cite{Harshath2021} demonstrates an encouraging use of more advanced technologies like deep neural networks (DNN), random forest (RF), and linear discriminant analysis on classifying the land as farming/non-farming using geospatial information such as soil type, strength, climate, and type of crop.

To tackle the crop yield prediction problem, Fan et al.~\cite{Fan2022} proposes a combination of GNN and recurrent neural networks (GNN-RNN) approach incorporating geographic and temporal information. Similar to our work, they compare machine learning techniques trained on geographic factors and predict nationwide. They posit that GNN can boost the prediction power of a county's crop yield by combining the features from neighboring counties. Their result shows that the graph-based models (GNN and GNN-RNN) outperform competing baselines such as long short-term memory (LSTM) and convolutional neural network (CNN), illustrating the importance of geographic context in graphs.

\subsection{Landmine Detection}
The research that focuses on landmine detection problems with machine learning can be categorized into two groups according to different input sources. The first group of methods reads remote sensing data such as satellite images, hyperspectral images, or normalized difference vegetation index (NDVI). Several research demonstrated the usefulness of image data~\cite{Makki2017,Alzoubi2021}. Still, according to the Makki et al.~\cite{Makki2017}, the detection performance suffers from a trade-off between computational complexity and detection performance. Furthermore, different types of remote sensing produce varying advantages in different environments. Therefore, the benefit of using remote sensing is highly dependent on the use case~\cite{Jebens2021}.
Also, it is challenging to directly correlate landmine risk with the environmental factors that impact them from the remote images. As a result, another approach focuses on gathering ecological factors and using them as inputs to train models directly.

While early works such as~\cite{Alegria2011,Chamberlayne2002} mainly focus on spatial statistics analysis in combination with GIS usage,~\cite{Rafique2019} and~\cite{Schultz2016} could be considered highly relevant for our work because of their similarities in implementing ML techniques on landmine risk prediction problems. They also use explanatory variables from mainly open source data, including land cover (water channel and buildings), remnants of war indicators (control area, conflict area, medical facility, roads, and border), and topography (elevation and slope). Schultz et al.~\cite{Schultz2016} studies the presence of landmines in different scales of areas however without exploring granularity aspects that are important for geographical and political considerations. 
Rafique et al.~\cite{Rafique2019} analyzes the information of confirmed hazardous areas and build a model to assess the risk of areas in the immediate vicinity of the hazardous areas. The prediction is progressively extended to farther areas once the suspect is confirmed or discarded. However, this approach slows down the desk assessment step. The conservative approach of Rafique et al.~\cite{Rafique2019} is motivated by the natural ambiguity in the data since while a CHAs is surely hazardous, areas outside the CHAs are often uncharted and therefore might be either hazardous or safe.
This makes the ML models to reach good performance but limits the use case to predicting only in reduced areas. There exist a recent follow-up research which applies the Desk-Aid approach for the mines in Colombia~\cite{rubio2023reland}. The study also focuses on the interpretable models and domain expert usage of the system.

\section{Desk-AId pipeline overview}

\begin{figure}[h]
 \centering
  \includegraphics[width=\linewidth]{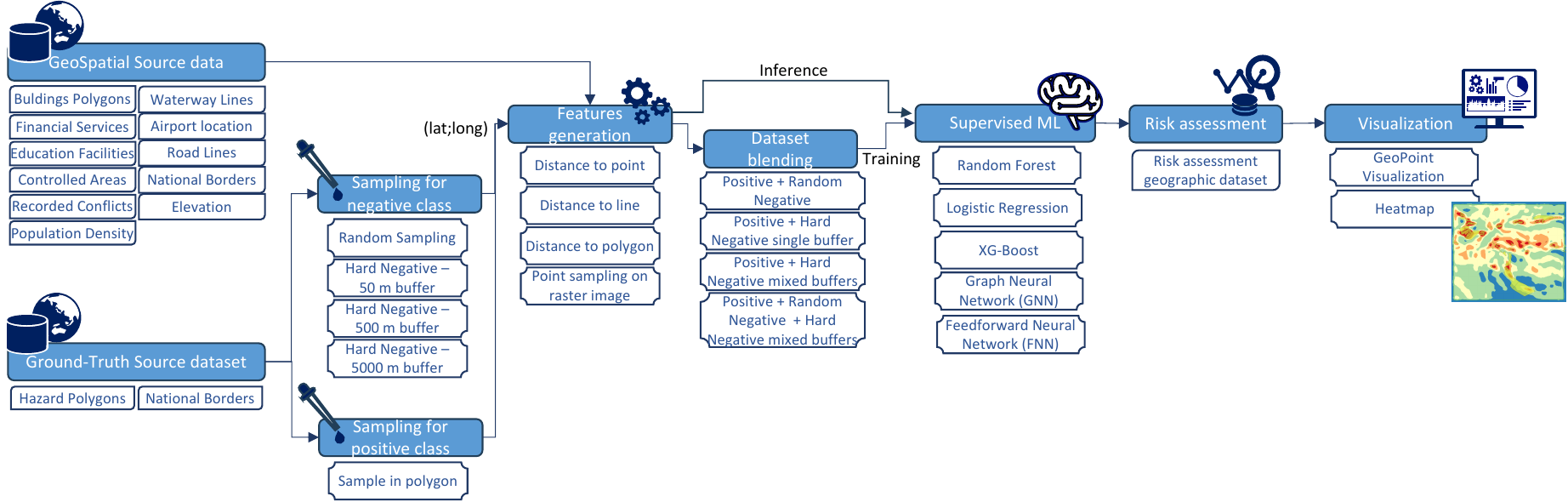}
  \caption{Desk-AId pipeline overview from data sources till risk assessment}
  \label{fig:pipeline_overview}
\end{figure}

Desk-AId is a data processing pipeline that involves multiple steps from the data acquisition till the AI-based risk assessment to support the landmine desk assessment. Fig.~\ref{fig:pipeline_overview} shows the overall pipeline with details of the different approaches and techniques we have adopted in this article. First, the geospatial source data is acquired from multiple sources. More information on the data is report in Table~\ref{table:dataoverview}. Then, we apply different sampling techniques for positive points (i.e., within hazardous areas) and for negative points. In the first case, we have a ground-truth datasets of polygons representing confirmed hazardous areas. Thus, we sample points in the form of (latitude;longitude) within the polygons. In the second case, we do not have confirmed clear areas, therefore, we sample geographical points within the national borders excluding the hazardous polygons. This sampling generates noisy data since it is not sure whether the sampled point is actually a negative point. We adopt different techniques to mitigate the noise that are presented later in section~\ref{sec:dataset}.

The pipeline, then, uses the location of the samples to sample the geospatial data sources and generate features using different algorithms. In this article, we use state-of-the-art algorithms implemented in the QGIS framework. After this step, we have location and features for negative and positive samples generated with different approaches. We blend these samples to compile a balanced dataset for training a machine learning model. We experiments as ML models classifiers and more sophisticated neural networks. In particular, we model a graph using geographic distance train a graph neural network (GNN) (more details in section~\ref{sec:ml})

The trained model are applied to any geographic point (generated either by the sampling for the ML testing, or selected by the domain expert for the desk assessment) to infer risk assessment that can be visualized into a geographic system.
\section{Building the dataset}
\label{sec:dataset}

We build the dataset addressing two challenges: i) use open available data for maximize solution replicability, ii) deal with the lack of ground-truth for negative points.
The data considered for the ground-truth is a collection of CHAs polygons and a set of geographical and socio-economic datasets. All the dataset files are in the format of the shapefile (shp) and Geographic Tagged Image File Format (tiff). We utilize the open-source software that is used as a platform by the domain expert:  Quantum Geographic Information System (QGIS) to process the datasets, calculate and generate features, and output the aggregated data as CSV files.

\begin{figure}[h!]
\centering
\begin{minipage}{.63\textwidth}
  \centering
	\includegraphics[width=\linewidth]{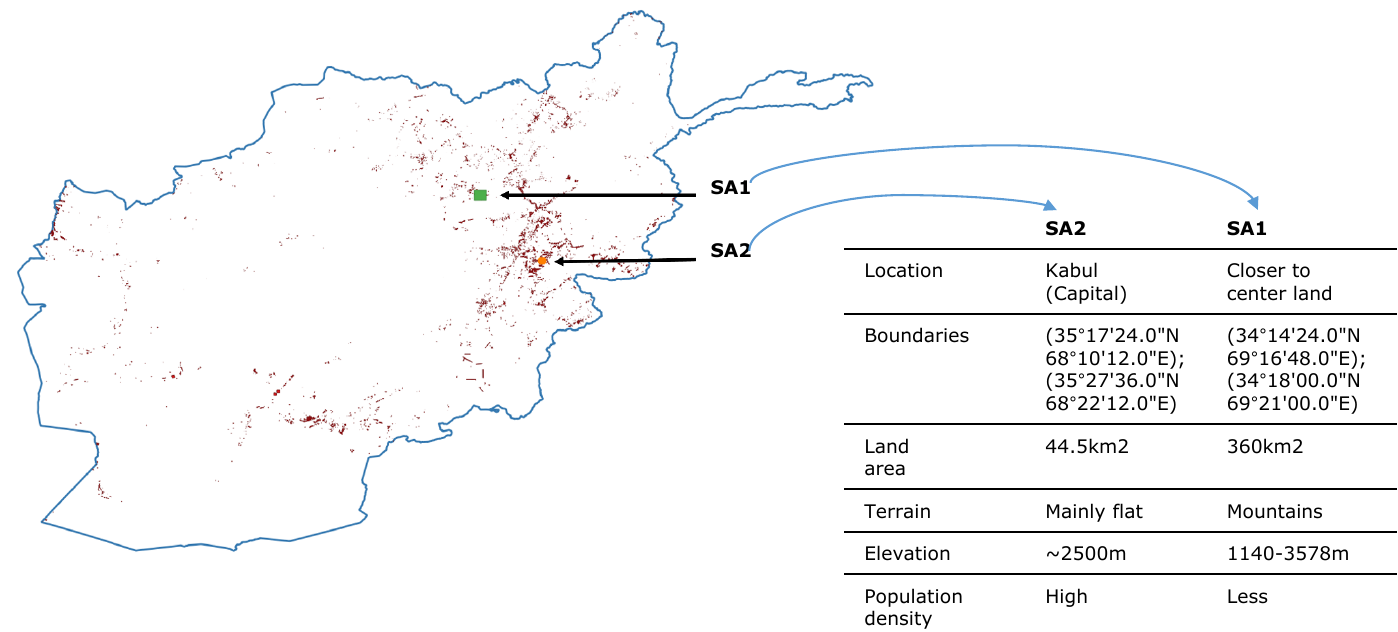}
	\caption[Landmine distribution and the two study areas in Afghanistan country land]{Landmine distribution and the two study areas in Afghanistan country land. Green area is the study area one (SA1) and orange one denotes study area two (SA2) that is around Kabul. Minefields are marked in red. }
	\label{fig:SA12}
\end{minipage}%
\hspace{0.75em}
\begin{minipage}{.32\textwidth}
  \centering
  \includegraphics[width=\linewidth]{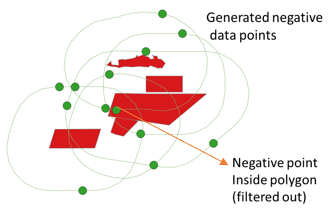}
  \caption{Hard negative samples generated around the positive samples (landmine polygons colored in red).}
  \label{fig:hn_generation}
\end{minipage}
\end{figure}

\subsection{Source data to features}

We build the dataset starting from the hazard polygons reported by Afghan mine action NGOs and authorities before 2020. The recorded hazard data has been collected originally by numerous NGOs and authorities for decades and entered into the Information Management System for Mine Action (IMSMA) system\footnote{IMSMA Wiki, \url{https://mwiki.gichd.org/IM/Main_Page}}. From this data, we select the relevant hazard types, such as landmine and explosive remnants of war (ERW). 
The data is formed by 12,098 polygons spread throughout Afghanistan (see the red areas in Fig.~\ref{fig:SA12}). The hazard polygons dataset is containing also other variables that we do not take into consideration since we cannot reproduce for the negative points.

We, then, collect geographical data from multiple sources covering different type of information spread within the national border of Afghanistan. For the selection of the features, we aim at the most abundant data in order to easily apply the system into another country or region. We include dataset reporting building presence with polygons, financial infrastructure, education facilities, health facilities, road lines and water ways. From these information, given the coordinates of a data point, we calculate the distance to the closest polygon or point (depending on the dataset). We utilize a QGIS package 'Distance to nearest hub (points)'. The algorithm computes the distance between the origin features of sample points and their closest destination. Distance calculations are based on the center of the feature, and the resulting layer contains the origin features center point with an additional field indicating the identifier of the nearest destination feature (the categorical feature here) and the distance to it. For example, we input the sample points as the source points layer and indicate education facilities as the destination hubs layer. The algorithm outputs the distance between the sample points and their nearest education facility.  
Depending on the meaningfulness of the original dataset (sometimes we face high portion of null data), we also report as feature the type of entity such as school or university for education facilities, or clinic or hospital for health facilities. The number of categories for each features varies between 4 for education facilities, 8 for air traffic facility, 6 plus 1 for unknown category for health facility, and 10 for controlled area authority.

Calculating the distance of a sample point to a line is similar to calculating the distance to the nearest point. However, the line essentially differs from the point in that it's naturally hard to define the a center, which the QGIS package calculated from. Indeed, the distance could not be correctly calculated when setting the line as a destination layer. Therefore, a way to workaround is to first transfer a line to multiple points and then use the 'Distance to nearest hub (points)' package. We utilize a built-in algorithm "Extract Vertices", which takes a line as input and generates a point layer with points representing the vertices in the input line. The road lines are transferred to 4,649,404 points, and the waterway lines are transferred to 968,161 points. Then, each points layer is served as a destination that the interest points calculate the nearest distance. 

We include also a Georeferenced Event Dataset dataset from the Uppsala Conflict Data Program (UCDP)~\cite{UCDP}. This dataset reports events of organized violence with at least one casualty. We use this dataset to calculate distance to a conflict and estimate death for each data point. 

Finally we take into consideration also population density information and topographical information such as elevation.
Deriving population density and elevation for interest points fundamentally differ from generating features from polygons, points, or lines since the sources are raster data with continuous data values all over the country. To extract the value for interest points, we utilize the QGIS plugin "Point sampling tool"\footnote{Point sampling tool, \url{https://github.com/borysiasty/pointsamplingtool}}. This tool samples raster values from all the features of the raster cells given a geographical point. Thus, we can specify the sample points and make the algorithm create a file containing population density and elevation value at the location of the sample point. 
The hill slope in percentage data is calculated from the elevation layer using the 'slope' package from GDAL\footnote{GDAL documentation,\url{https://gdal.org/programs/gdaldem.html}} in QGIS, the output of which is also a 30-meter grid raster layer, similar to elevation. Since it is raster data, we also utilize the point sampling tool to extract the value at the location of interest points.

\begin{table}[]
\resizebox{\textwidth}{!}{\begin{tabular}{llrlll}
\hline
\hline
\textbf{Data}                                & \textbf{Format}       & \textbf{Entries}        & \textbf{Data source}                                                      & \textbf{Features Generated}                    & \textbf{Type of data}                                                                                                      \\ \hline
Hazard Polygons                              & shp                   & 12,098                  & \begin{tabular}[c]{@{}l@{}}Afghan mine\\action authorities\end{tabular}                                            & Hazard presence                                & Categorical (i.e., 1)                                                                                                      \\
\hline
Building Polygons                            & shp                   & 1,429,520               & OpenStreetMap \cite{OpenStreetMap}                  & *Distance to closest building (m)               & Continuous                                                                                                                 \\
\hline
Financial Services                    & shp                   & 138                     & OpenStreetMap \cite{OpenStreetMap}                  & Distance to closest financial service (m)      & Continuous                                                                                                                 \\
\hline
\multirow{2}{*}{\begin{tabular}[c]{@{}l@{}}Education\\Facilities\end{tabular}} & \multirow{2}{*}{shp}  & \multirow{2}{*}{466}    & \multirow{2}{*}{OpenStreetMap \cite{OpenStreetMap}} & Distance to closest education facility (m)     & Continuous                                                                                                                 \\
                                             &                       &                         &                                                                           & Closest education facility type                & \begin{tabular}[c]{@{}l@{}}Categorical {[}school, college \\ university, kindergarten {]}\end{tabular} \\ 
\hline
\multirow{2}{*}{Airport}              & \multirow{2}{*}{shp}  & \multirow{2}{*}{230}    & \multirow{2}{*}{OpenStreetMap \cite{OpenStreetMap}} & Distance to closest air traffic facility (m)   & Continuous                                                                                                                 \\
                                             &                       &                         &                                                                           & Closest air traffic facility type              & \begin{tabular}[c]{@{}l@{}}Categorical {[}helipad, aerodrome \\ gate, etc. {]}\end{tabular} \\
\hline
\multirow{2}{*}{Health Facilities}    & \multirow{2}{*}{shp}  & \multirow{2}{*}{231}    & \multirow{2}{*}{OpenStreetMap \cite{OpenStreetMap}} & Distance to closest health facility (m)        & Continuous                                                                                                                 \\
                                             &                       &                         &                                                                           & Closest health facility type                   & \begin{tabular}[c]{@{}l@{}}Categorical {[}clinic, hospital, \\ ..., unknown {]}\end{tabular} \\
\hline
Road Lines                                   & shp                   & 191,531                 & OpenStreetMap \cite{OpenStreetMap}                  & *Distance to closest road (m)                   & Continuous                                                                                                                 \\
\hline
Waterway Lines                               & shp                   & 16,948                  & OpenStreetMap \cite{OpenStreetMap}                  & *Distance to closest water way (m)              & Continuous                                                                                                                 \\
\hline
\multirow{2}{*}{Controlled Area}         & \multirow{2}{*}{shp}  & \multirow{2}{*}{44,400} & \multirow{2}{*}{OpenStreetMap \cite{OpenStreetMap}} & \begin{tabular}[c]{@{}l@{}}Distance to closest area\\under authority control (m)\end{tabular}     & Continuous                                                                                                                 \\
                                             &                       &                         &                                                                           & Authority of the controlled area               & \begin{tabular}[c]{@{}l@{}}Categorical {[}gov. (before 2020), \\ local Taliban, Jamiat Islami, etc. {]}\end{tabular} \\
\hline
\multirow{2}{*}{\begin{tabular}[c]{@{}l@{}}Recorded\\Conflicts\end{tabular}}        & \multirow{2}{*}{shp}  & \multirow{2}{*}{41,452} & \multirow{2}{*}{\begin{tabular}[c]{@{}l@{}}UCDP\\\cite{UCDP}\end{tabular}}                   & \begin{tabular}[c]{@{}l@{}}Distance to closest recorded\\conflict event(m)\end{tabular} & Continuous                                                                                                                 \\
                                             &                       &                         &                                                                           & Estimated death (casualties)                   & Continuous                                                                                                                 \\
\hline
\begin{tabular}[c]{@{}l@{}}National Borders\\Line\end{tabular}                         & shp                   & 1                       & \begin{tabular}[c]{@{}l@{}}GeoBoundaries\\\cite{geoBoundaries}\end{tabular} & *Distance to border (m)                         & Continuous                                                                                                                 \\
\hline
Population Density                           & tiff                  & NA                      & Grid Population \cite{NASA}                         & *Population Density (persons/km$^{2}$)            & Continuous                                                                                                                 \\
\hline
\multirow{2}{*}{Elevation}                   & \multirow{2}{*}{tiff} & \multirow{2}{*}{NA}     & \multirow{2}{*}{\begin{tabular}[c]{@{}l@{}}ASTER GDEM\\\cite{ASTER}\end{tabular}}            & *Elevation (m)                                  & Continuous                                                                                                                 \\
                                             &                       &                         &                                                                           & *Hill Slope (\%)                                 & Continuous                                                                                                                 \\ \hline
\end{tabular}}
\caption{Original used datasets overview and respective features generated. The features marked with a star (*) are the base set of features used in all the ML models.}
\label{table:dataoverview}
\end{table}

\subsection{Dataset for ML training}

We build a balanced dataset for the ML training from sampling points within polygons with landmine presence (reported in the hazard dataset, see Table~\ref{table:dataoverview}) and absence regions, respectively. 
We sample two points for each hazard polygon regardless to the size of the polygon in order to avoid the information loss of small polygons. 
After this step, we have the location of the positive data points, and we can start mapping points with explanatory variables calculated from the geographical layers (Table~\ref{table:dataoverview}).

Then, we collect the same number of data point in the hazard absence class that are the negative points. The first approach we follow is to randomly sample points~\cite{Schultz2016} throughout the country of Afghanistan that do not lay within the hazard polygons. As seen in Fig.~\ref{fig:SA12}, the data is very sparse and some areas are very far from the recorded hazard. That does not mean that there are no hazard in those locations, since we are only sure on the hazard presence for areas covered in the given dataset while we do not have sure information for other geographic areas.

As second approach for the negative sampling, we exploit the concept of hard negative mining. We define a buffer zone around the hazard polygons using a heuristic distance, and select a point within such area. In this way we ensure that the negative samples have higher similarity to the positive sample. 
We build the negative samples using three buffer zone distances, namely 50 meters, 500 meters, and 5000 meters. The numbers are chosen heuristically from the observation that the minimum distances from features to sample points (e.g., \textit{Distance to Building}) are roughly 50 meters. Therefore, the three distances are chosen to experiment with the effect of buffers. Fig.~\ref{fig:hn_generation} illustrates an example of generation of hard negative points in the buffer zones of hazard polygons. The sample data points in a buffer zone that are inside another hazard polygon is filtered out to guarantee an equal number of negative samples.
We utilize a QGIS "Buffer" tool to draw the buffer zones, change polygons to the line, and assign points on the buffer line using QGIS "QChainage" plugin\footnote{QGIS plugin "QChainage", \url{https://github.com/mach0/qchainage}}. 

\begin{figure}[t]
	\begin{center}
	\includegraphics[width=0.8\linewidth]{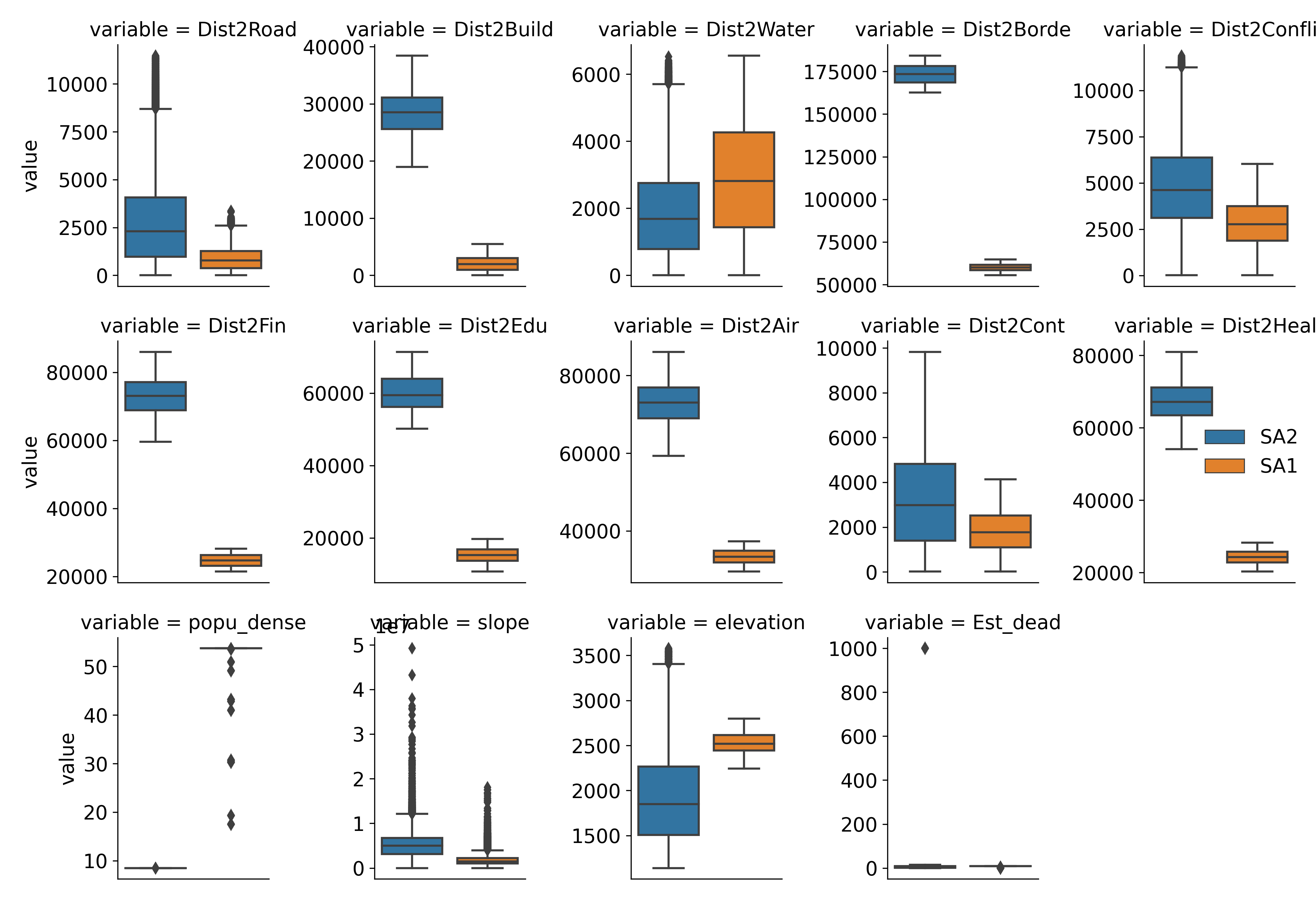}
	\caption{Box plot of numeric features of the two study areas SA1 and SA2.}
	\label{fig:box_plot}
	\end{center}
\end{figure}

\subsection{Study Areas}

In addition to the desk assessment of the whole country, we also explore the performances of our approach when applying an ML model to data points laying into an unseen areas. For this purpose we select two study areas with different characteristics. The two chosen study areas are also in the Fig.~\ref{fig:SA12}. 
Study area 1 (SA1) is located closer to the center of the land and has a very low population density (around 8.5 persons/km$^{2}$). It covers the land of 360 square kilometers. One obvious distinction of SA1 is the variety of elevation, which ranges from 1140 to 3578 meters. The slope ranges from zero to five. 
Study area 2 (SA2) is close to the capital of Afghanistan, Kabul (34.31 N, 69.12 E). It covers the land of 44.5 square kilometers. The terrain in this region is mainly flat, with an elevation of around 2500 meters. Because the area is near the capital, it has a relatively high population density (ranges between 17.5 and 53.7 persons/km$^{2}$) and a low distance to the community facilities such as roads, buildings, financial and educational facilities, or airport and health sites. 
Because of the terrain and size of study areas, the features of SA1 have a wider variety than SA2. 
Fig.~\ref{fig:box_plot} illustrates a detailed comparison of numeric features in both study areas.

\section{Machine learning models}
\label{sec:ml}

We test the generated datasets with classifiers and with neural networks. We use two feature sets to examine the effect of adding attributes. One set is formed of seven features (marked with a star in Table~\ref{table:dataoverview}) and the other is the expanded set with eighteen features. 
The neural network models use always all eighteen attributes since they are more robust on abundant feature set.


We first implement a feedforward artificial neural network (FNN) and we choose the standard optimizer AdamW~\cite{Loshchilov2018}. An early stopping function from the Keras package is set to avoid overfitting. Since the dataset is balanced in this case, we choose the validation accuracy (\textit{val\_accuracy}) as the monitor and set patience as 50. 

Further, we explore the Graph Convolutional Neural Networks (GCNNs) as it utilizes the graph structure to gather node information from neighborhoods in a convolutional manner. The main idea of GCNNs is to generate a node $v$'s representation by aggregating its own features $x_v$ and neighbors' features $x_u$, where $u \in N(v)$. It has been proven to be well-suited for modeling a graph consisting of interconnected geographic locations~\cite{Zhu2020}. 

Before implementing GCNNs, we need to build a graph. Considering the characteristic of GCNNs and the available data, we define a location-based graph structure as follows: 
assuming a set of location (points on the map) where each location point has characteristic $\mathbf{X}$ that can be represented as feature vectors [$x_1$, $x_2$...] and $x_i$ denotes the values for the $i$th dimension of $\mathbf{X}$. $E$ refers to the connection between the location points. Considering the complexity and the purpose of this work, we use the QGIS package 'Distance Matrix' to identify the five nearest neighbors and calculate the distance to each point. Then, a location-based graph $G = (V,E)$ is constructed to connect location points as a graph. Each point on the map can be formalized as a node $v_i \in V$ in $G$ and the point features $\mathbf{X}$ are encoded as the node attributes $x_k \in \mathbf{X}$ on every $v_k \in V$. 
On the other hand, the place connection is represented as the edge $E$ where $e_{i j}=\left(v_{i}, v_{j}\right) \in E$ denote an edge pointing from $v_{i}$ to $v_{j}$. As stated before, the neighborhood of a node $v$ is defined as $N(v)=\{u \in V \mid(v, u) \in E\}$. Here, we have the edge attributes $\mathbf{X}^e$ where $\mathbf{X}^{e} \in \mathbf{R}^{m \times c}$ is an edge feature matrix with $\mathbf{x}_{v, u}^{e} \in \mathbf{R}^{c}$ representing the feature vector of an edge, i.e., the distance between the two location points $v_{i}, v_{j}$. After the graph is defined, it is ready to be implemented in GCNNs where it generates node $v$'s class (i.e. presence of landmine) by aggregating $v$'s own features $x_v$ and neighbors' features $x_u$, where $u \in N(v)$. 

To load the graph data into the model, we aggregate the graph information into a tuple:

\begin{enumerate}
    \item node features $x_k$: a two-dimensional where each row corresponds to a node and each column corresponds to a feature of that node. 
    
    \item edges $e_{i j}$: a shape of a two-dimensional array with two rows and a number of columns equal to the number of edges. The first row corresponds to the starting node of an edge and the second row corresponds to the ending node of an edge. We take the links between the five nearest neighbor points. 
    
    \item edge weights $\mathbf{x}_{v, u}^{e}$: a one-dimensional array with a length equal to a number of edges. It quantifies the relationships between nodes in the graph. The weight corresponds to the distance between two location points. 
\end{enumerate}

We implement the GCNNs node classification model following the approach from You et al.~\cite{You2020}. First, we apply the FNN module that we have implemented for preprocessing the node features to generate initial node representations. 
Next, we choose to implement two layers of graph convolution built from the graph information to build the node embeddings. Too many graph convolutional layers may cause the problem of oversmoothing~\cite{Chen2020}. Finally, the FNN module is applied again to generate the final node embeddings, which are fed into a sigmoid layer to predict the node class for our binary classification problem.

\section{Experimental Result Analysis}
\label{sec:Experimental_Result_Analysis}

\subsection{Evaluating the approaches for negative sampling}

We implement our baseline pipeline considering the random sampling approach~\cite{Schultz2016} where positive samples (i.e., hazard presence) come from the records and negative samples are randomly sampled in the whole geographic area. In our case, we sample the data throughout the whole Afghanistan.

\begin{figure}[h!]
\centering
\begin{minipage}{.45\textwidth}
  \centering
  \includegraphics[width=\linewidth]{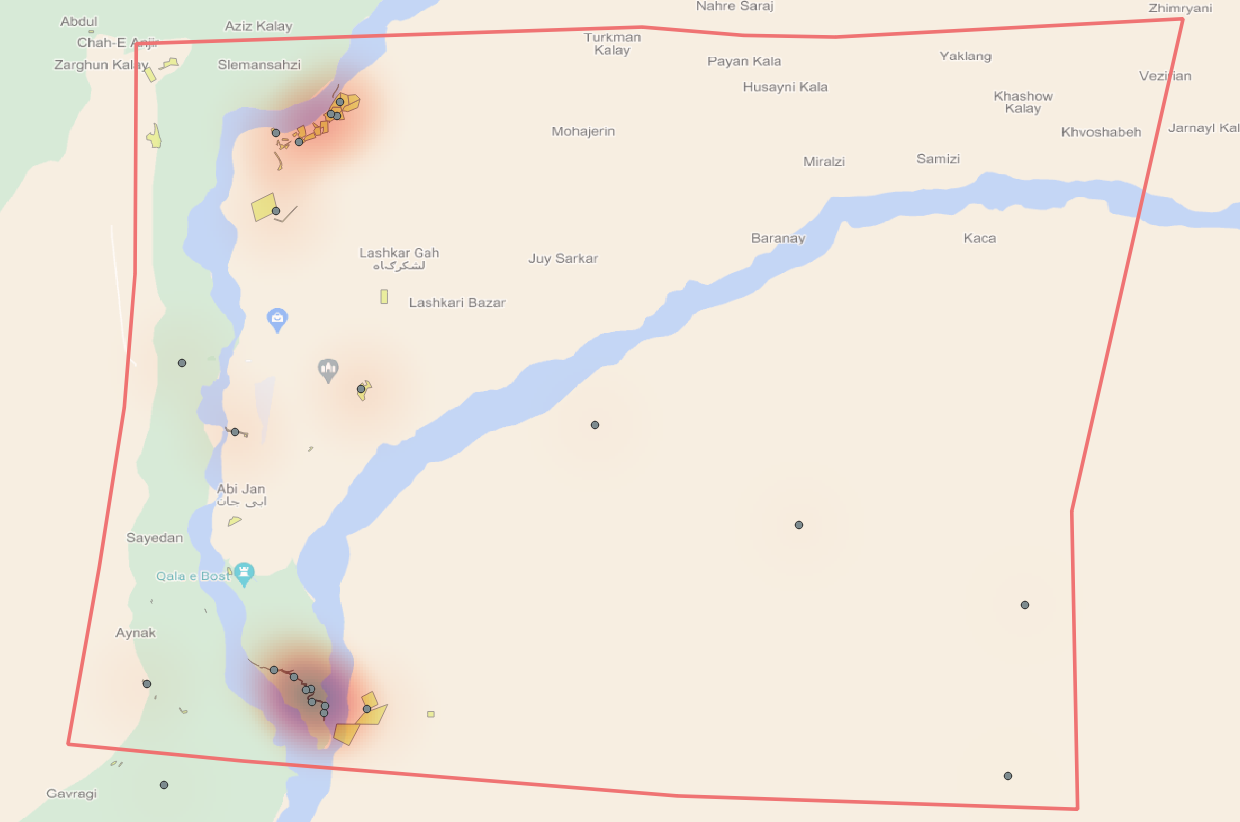}
  \caption{Visualization of the risk assessment. Dots are the testing samples, red areas are predicted as hazardous, and yellow polygons are ground-truth positive samples.}
  \label{fig:demo}
\end{minipage}
\hspace{0.5cm}
\begin{minipage}{.45\textwidth}
  \centering
  \includegraphics[width=\linewidth]{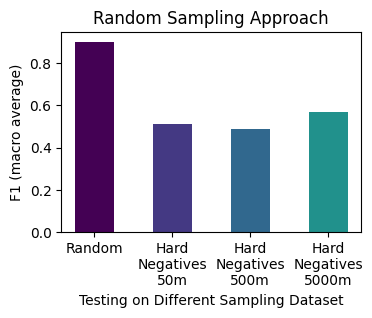}
  \caption{Testing different datasets when training the ML model with random sampling approach.}
  \label{fig:randomapproach}
\end{minipage}%
\end{figure}

We have trained three classifiers (i.e., Logistic Regression, Random Forest, and XGBoost) splitting the data in 75\% for training and 25\% of testing with a balanced dataset between the two classes. Random Forest is the classifier that reaches the best performance. This model achieves good performances with F1 score macro-averaged on the two classes equal to 0.9. We macro-average the F1 since the problem of desk assessment of mine detection requires correct classification of hazards (for human safety reasons) but also not misclassification of non-hazardous areas (for socio-economical reasons).
Fig. \ref{fig:demo} visualizes the results from the random sampling approach in a map. The demo area is in a province called Hilmand (31.36 N, 63.95 E), located in the southwest of Afghanistan. The distance between the positive and negative points ranges from 4100 to 28300 meters. In such scenarios, it is relatively easy for the ML model to distinguish between hazard areas and clear areas. 
For the desk assessment application, also it is important to reduce the sizes of sub-regions so as to reduce the effort (and costs) for the following steps in the demining operations. Therefore, we have tested the performance of this model against points that are close 50, 500 and 5000 meters to the recorded hazard. Fig.~\ref{fig:randomapproach} shows the results of the model trained on random sampling approach with different testing dataset. We can see that the performance when testing on points closer to the hazardous points drops between 0.49 to 0.57.




\begin{figure}[h]
  \centering
  \includegraphics[width=\linewidth]{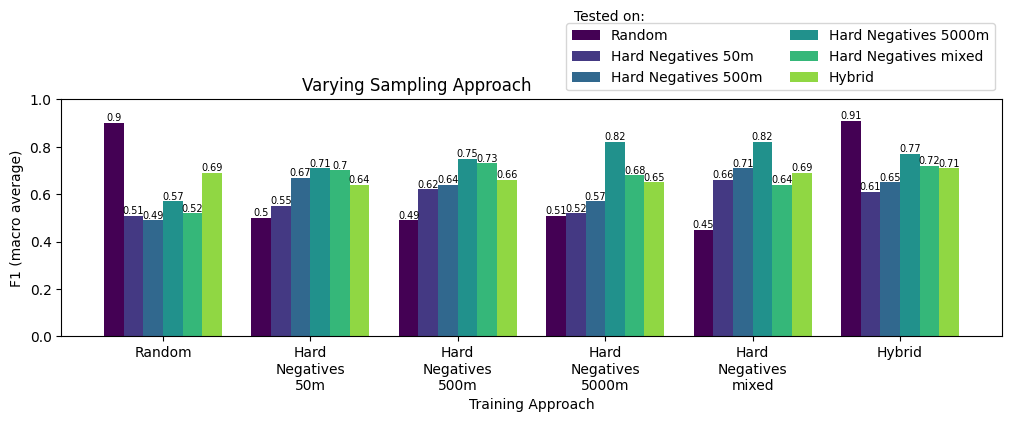}
  \caption{Evaluation of different negative points sampling approaches. Each island of bars refers to a different train dataset, each bar in the island refers to a different testing dataset.}  \label{fig:random_vs_hardnegatives}
\end{figure}


We, then, train the model by using sampling method with hard negatives considering three different buffer areas (50m, 500m and 5000m) and a balanced mix of them. Fig.~\ref{fig:random_vs_hardnegatives} shows the performances of the two approaches of random and hard negatives for the training for different testing dataset. We can see that the hard negatives approach outperform the random approach tested on the hard negatives. Nevertheless, we notice that when tested on the random dataset the hard negatives approaches do not perform well. This is due to the fact that the hard negative points are geographically linked to the positive points that are limited to particular geographical areas, where the confirmed hazardous areas (CHAs) are recorded. Therefore, the training models do not learn on locations for which we do not have any records such as the vast empty areas depicted in Fig.~\ref{fig:SA12}. Thus, we have tested a hybrid approach where we mix hard negatives of all the three buffer sizes with random negatives points to represent the rest of the country. The last set of bars in Fig.~\ref{fig:random_vs_hardnegatives} shows that this approach is the best trade-off for all the testing scenarios.

\subsection{Testing the Study Areas}
\label{sec:HS_SA}

\begin{figure}[h]
\centering
\begin{minipage}{.45\textwidth}
  \centering
  \includegraphics[width=\linewidth, height=4cm]{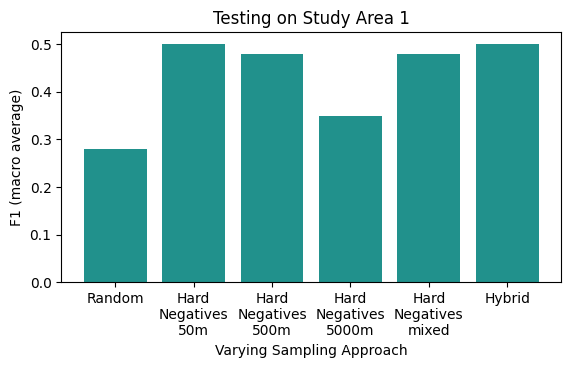}
  \caption{Testing different negative points sampling approach on study area 1 (SA1).}
  \label{fig:random_vs_hardnegatives_sa1}
\end{minipage}%
\hspace{0.5cm}
\begin{minipage}{.45\textwidth}
  \centering
  \includegraphics[width=\linewidth, height=4cm]{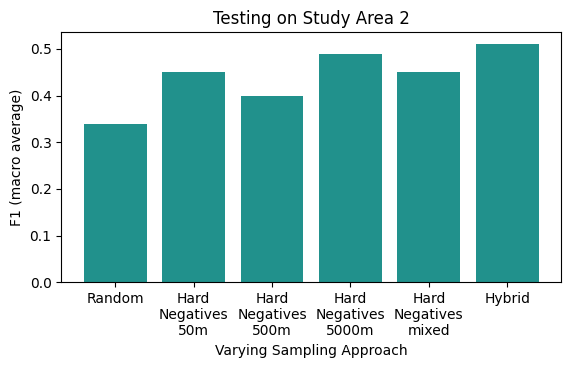}
  \caption{Testing different negative points sampling approach on study area 2 (SA2).}
  \label{fig:random_vs_hardnegatives_sa2}
\end{minipage}
\end{figure}

We test the different trained models also against the two study areas.  Notice that the landmine contamination in both study areas is highly imbalanced. SA1 has 9\% and SA2 has only 6\% of landmine presence.  Fig.~\ref{fig:random_vs_hardnegatives_sa1} and Fig.~\ref{fig:random_vs_hardnegatives_sa2} shows the results. For each experiment, we remove from the training dataset the points that lay within the study area under focus and test on the points within the study area. The hard negatives and hybrid approach are the approaches that reach the best results.


\begin{figure}[h]
\begin{minipage}[t]{0.43\textwidth}
\centering
\includegraphics[width=0.8\textwidth]{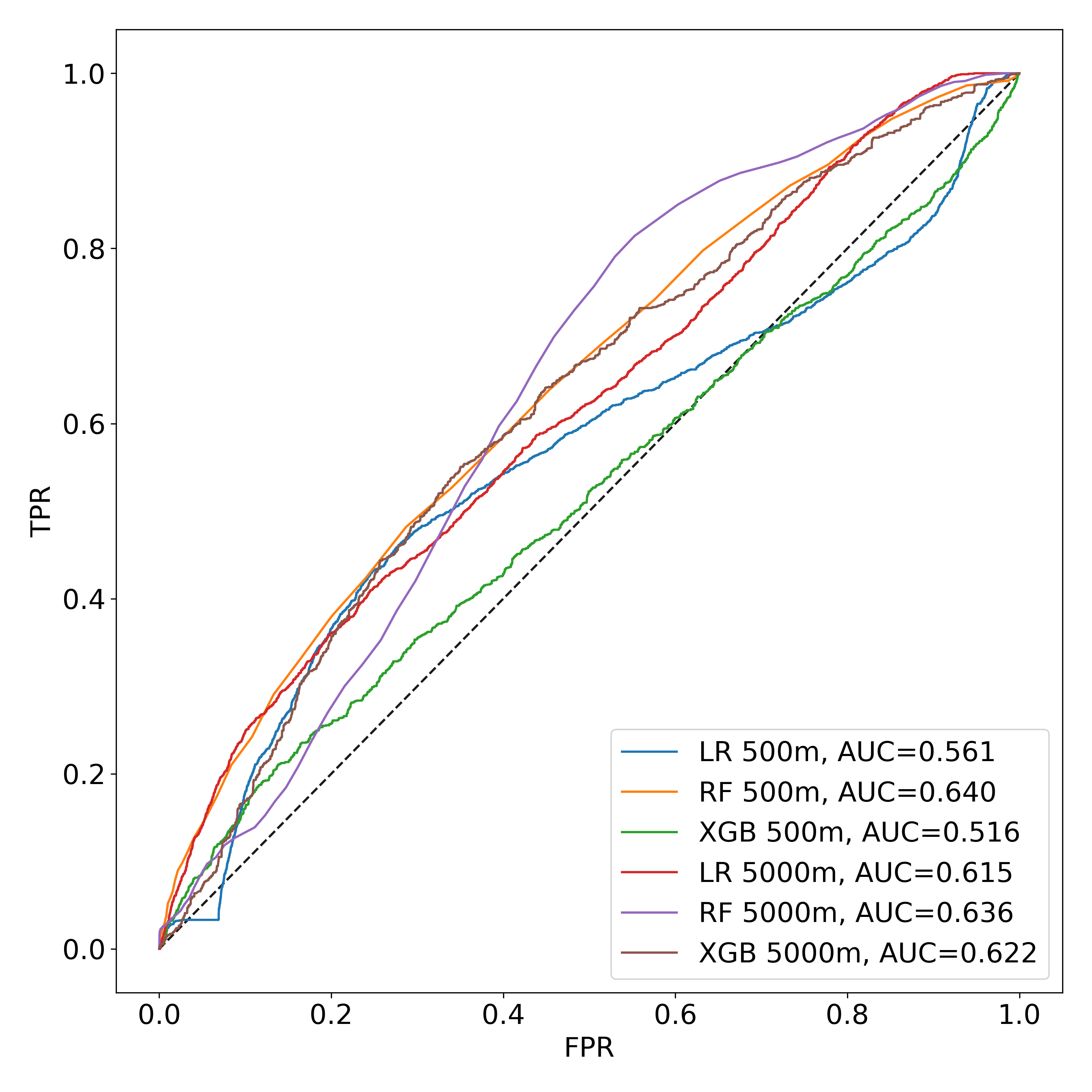}
\caption{ROC Curve for 500m and 5000m hard samples dataset with 18-features set tested on SA1.}
\label{fig:ROC_SA1_M_L}
\end{minipage}
\hfill
\begin{minipage}[t]{0.48\textwidth}
\centering
\includegraphics[width=0.8\textwidth]{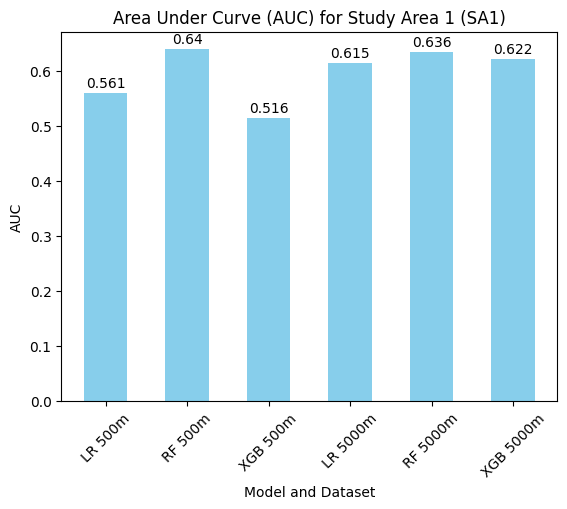}
\caption{AUC scores SA1 for 500m and 5000m hard samples dataset with 18-features set tested on SA1.}
\label{fig:AUC_SA1_M_L}
\end{minipage}
\end{figure}

Fig. ~\ref{fig:ROC_SA1_M_L} and Fig. ~\ref{fig:AUC_SA1_M_L} show the Receiver Operating Characteristic (ROC) curve and the area under the ROC curve (AUC). The ROC curve is a graphical representation of the trade-off between true positive rate (sensitivity) and false positive rate (1 - specificity) across different threshold values. For landmine risk assessment task, this translates as the capability of the model to correctly identify landmines over all the predicted positives versus the tendency of the model to incorrectly tagging a point as positive among all the predicted positives.
AUC measure the area under the ROC curve. Fig. ~\ref{fig:AUC_SA1_M_L} shows that RF outperforms other models with the highest AUC score for 500m (0.640) and 5000m hard negative samples (0.636). XGB performs the second best for the largest buffer (AUC = 0.622). 

To understand the difference in the two study areas, we plot the numeric feature distribution for both areas as box plots in Figure \ref{fig:box_plot}. We can observe that most of the features in SA2 have a more comprehensive range of values in the data distribution. The two study areas are selected in two regions with different characteristics: SA2, compared to SA1, is in the rural county with less population, large slope and elevation distribution, and high distance variability to points/polygons. 
This characteristic of data gives a high potential for RF to distinguish the testing data points as it was trained from the whole country's land and has covered a wide range of feature distribution.

The high feature variability in SA2 could also be used to explain the poorer performance of XGB and LR. The box plots in Figure \ref{fig:box_plot} visualize outliers as a data point located outside the box plot's whiskers. There are outliers in features such as \textit{Distance to Road}, \textit{Distance to Water}, \textit{Distance to Conflict Area}, \textit{Slope}, and \textit{Elevation} in SA2. 
XGB is known to be more sensitive than other tree-based models, such as RF, because its gradient boosting is easily impacted by outliers. When learning from the whole country land and taking a wide range of features into training, it has a high risk of being overfitted to the outlier. Same for LR, outliers could significantly influence the decision boundary. On the contrary, RF takes the average of multiple decision trees, reducing the impact of outliers.

\subsection{Adding features to the dataset}

\begin{figure}[h]
	\begin{center}
	\includegraphics[width=0.7\linewidth]{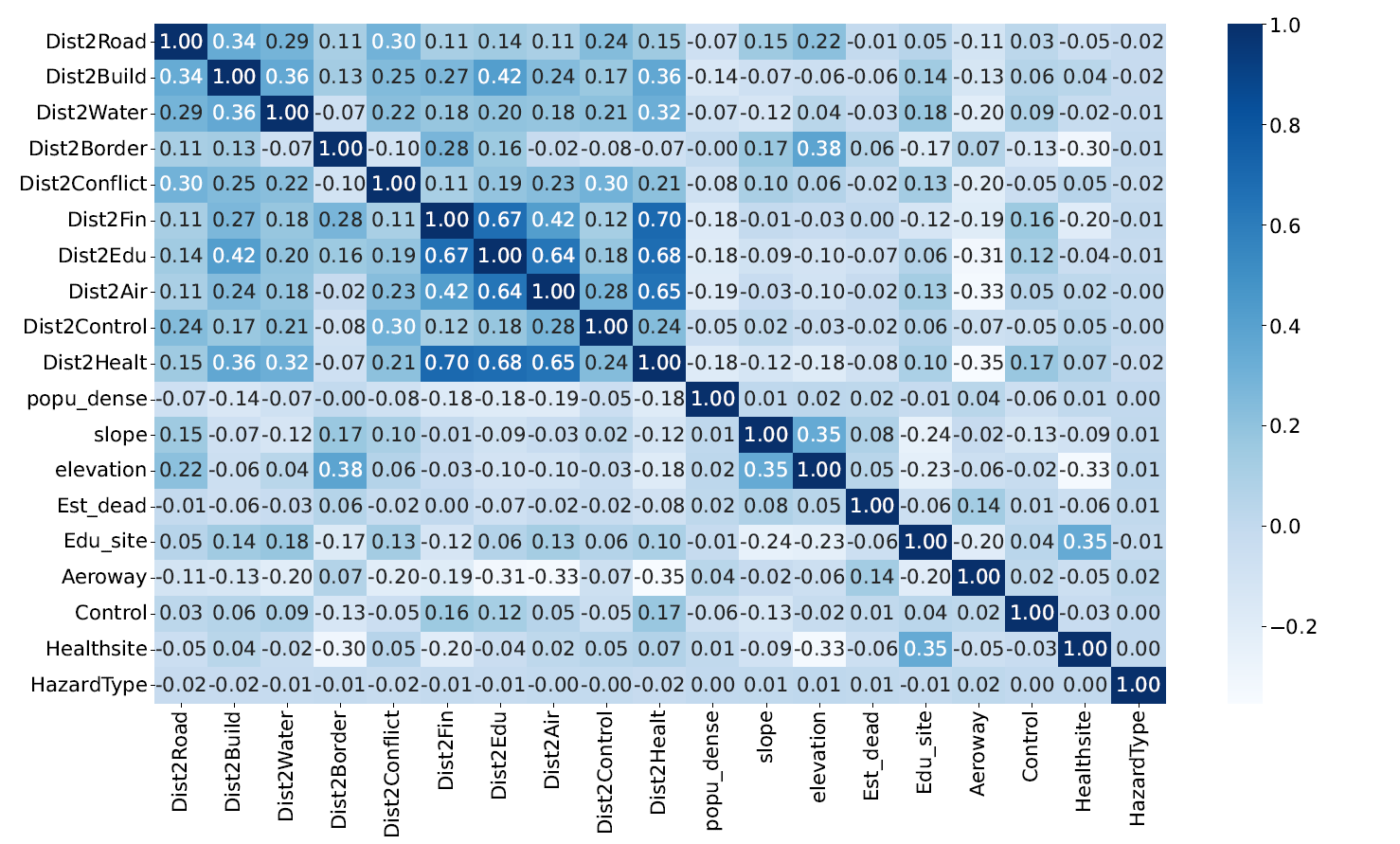} %
	\caption[Feature Correlation Matrix of 500m Hard Samples]{Feature Correlation Matrix of 500m Hard Samples. 'Dist2' implies 'Distance to'. 'Fin' stands for 'Financial Service'. 'Edu' is 'Education Facility'. 'Air' means 'Airport'. The last four features are the categorical features.}
	\label{fig:correlation_matrix}
	\end{center}
\end{figure}

In order to improve the performance, we included all the 18 features for training the model. 
Fig. \ref{fig:correlation_matrix} shows the correlation matrix for the hard sampling approach configured with 500 meters buffer. 
We can notice hardly any feature with a high correlation with the target label \textit{HazardType} that indicates a landmine presence. Nevertheless, some features show noticeable relations with others.

\begin{figure}[h]
\centering
\begin{minipage}{.45\textwidth}
  \centering
  \includegraphics[width=\linewidth, height=5.5cm]{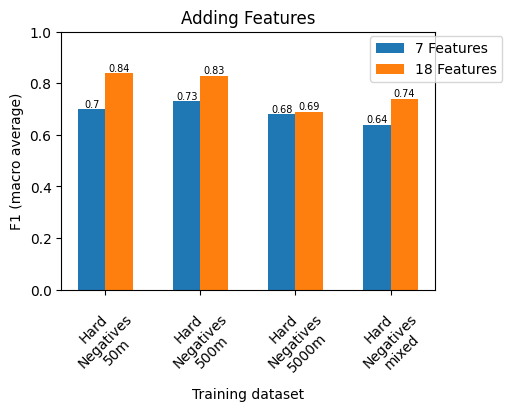}
  \caption{Comparing RF model performance with two features set: 7 features and 18 features.}
  \label{fig:featureset_comparison}
\end{minipage}
\hspace{.05\textwidth}
\begin{minipage}{.45\textwidth}
      \centering
      \includegraphics[width=\linewidth, height=5.5cm]{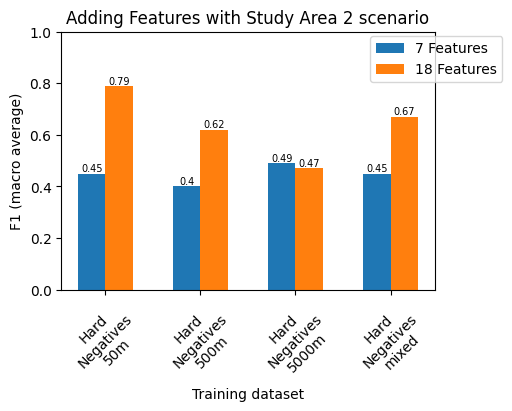}
      \caption{Comparing feature sets when testing on study area 2 (Kabul area) after training on rest of the country.}
      \label{fig:featureset_comparison_sa2}
\end{minipage}
\end{figure}


The effect of adding input features to the ML model can be compared in Fig. \ref{fig:featureset_comparison}. For all the experiments we have trained and tested again with that three classifiers noting that Random Forest is again performing best. 
We can observe that adding features is beneficial in all the scenarios, especially for the hard negatives with smallest buffer. An explanation of this might come to the fact that adding new features might lead to overfitting, reducing the capability of the classifier to generalize. This same phenomenon is noticeable when testing with the study area 2 scenario as depicted in Fig.~\ref{fig:featureset_comparison_sa2}.

The feature importance of 500m hard negative samples plotted in Table \ref{tab:feature_import_500m_7} and Table \ref{tab:feature_import_500m_18} also validate that some of the added features such as \textit{Distance to Control Area} and \textit{Distance to Conflict Area} are relevant for the model. Looking deeper into the feature importance, we observe that the reduced feature set have similar feature importance except for \textit{Population Density}. For the expanded feature set, the top important features mostly overlap with the reduced feature set, adding \textit{Distance to Control Area} and \textit{Distance to Conflict Area} on the list. 
We encounter similar behaviour with another tree-based model such as XGB. However, Linear Regression (LR) model does not show improvement.
This could be explained by calculating the Variance Inflation Factor (VIF)~\cite{Thompson2017} (see Table~\ref{tab:VIF}). 
A common rule of thumb is that a VIF score greater than 5 or 10 indicates high multicollinearity. 
The features added in the expanded feature sets such as \textit{Distance to Health Facility, Estimated Death, Distance to Airport} and \textit{Authority} may result in unstable coefficient estimates. This makes difficult to determine the true effect of each feature on the outcome.

\begin{table}[t]

    \parbox[t]{.30\linewidth}{
        \resizebox{\linewidth}{!}{
            \begin{tabular}{ |l|l| } 
             \hline
             Feature Name & VIF   \\ \hline
             Distance to Health Facility & 9.8303\\ 
             *Distance to Road & 9.6150 \\ 
             Estimated Death  & 9.0512 \\ 
             Distance to Airport & 6.5343 \\ 
             *Population Density & 6.4890 \\ 
             Authority & 6.3778 \\ 
             *Hill Slope & 5.7833\\ 
             \hline
            \end{tabular}
        }
        \vspace{1em}
        \caption{Features of VIF score larger than 5 from the expanded feature set in 500m hard negative samples. A star marks features in the base set.}
        \label{tab:VIF}
    }
    \hspace{0.5cm}
    \parbox[t]{.30\linewidth}{
        \resizebox{\linewidth}{!}{
            \begin{tabular}{|l|l|}
            \hline
            Feature Name & \\ \hline
            *Elevation & 0.1576\\
            *Distance to Border & 0.1575 \\
            *Distance to Waterway & 0.1562 \\
            *Distance to Building & 0.1548 \\
            *Distance to Road &0.1536 \\
            *Hill Slope & 0.1382 \\
            *Population Density &  0.0820\\
            \hline
            \end{tabular}
        }
        \vspace{1em}
        \caption{Sorted important features of the reduced feature set in 500m hard negative samples.}
        \label{tab:feature_import_500m_7}
    }
    \hspace{0.5cm}
    \parbox[t]{.30\linewidth}{
        \resizebox{\linewidth}{!}{
            \begin{tabular}{|l|l|}
            \hline
            Feature Name & \\ \hline
            *Distance to Road &0.0838 \\
            Distance to Control Area & 0.0819 \\
            *Distance to Waterway & 0.0806\\
            *Elevation & 0.0805 \\
            Distance to Conflict Area &0.0770 \\
            *Distance to Building & 0.0770 \\
            *Hill Slope &  0.0745\\
            \hline
            \end{tabular}
        }
        \vspace{1em}
        \caption{Top seven important features of the expanded feature set in 500m hard negative samples. A star marks features in the base set.}
        \label{tab:feature_import_500m_18}
    }
\end{table}

\subsection{Neural Network}

\begin{figure}[h]
\centering
\begin{minipage}{.323\textwidth}
  \centering
  \includegraphics[width=\linewidth]{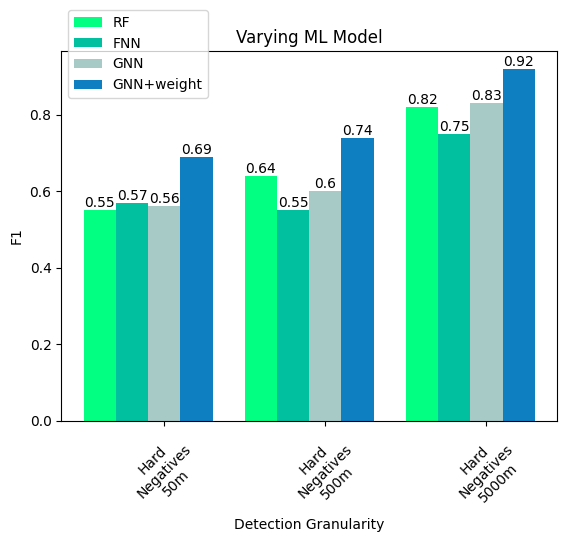}
  \caption{Comparing neural networks (NNs) models.}
  \label{fig:mlmodel_comparison}
\end{minipage}
\hspace{0.5cm}
\begin{minipage}{.62\textwidth}
  \centering
  \includegraphics[width=\linewidth]{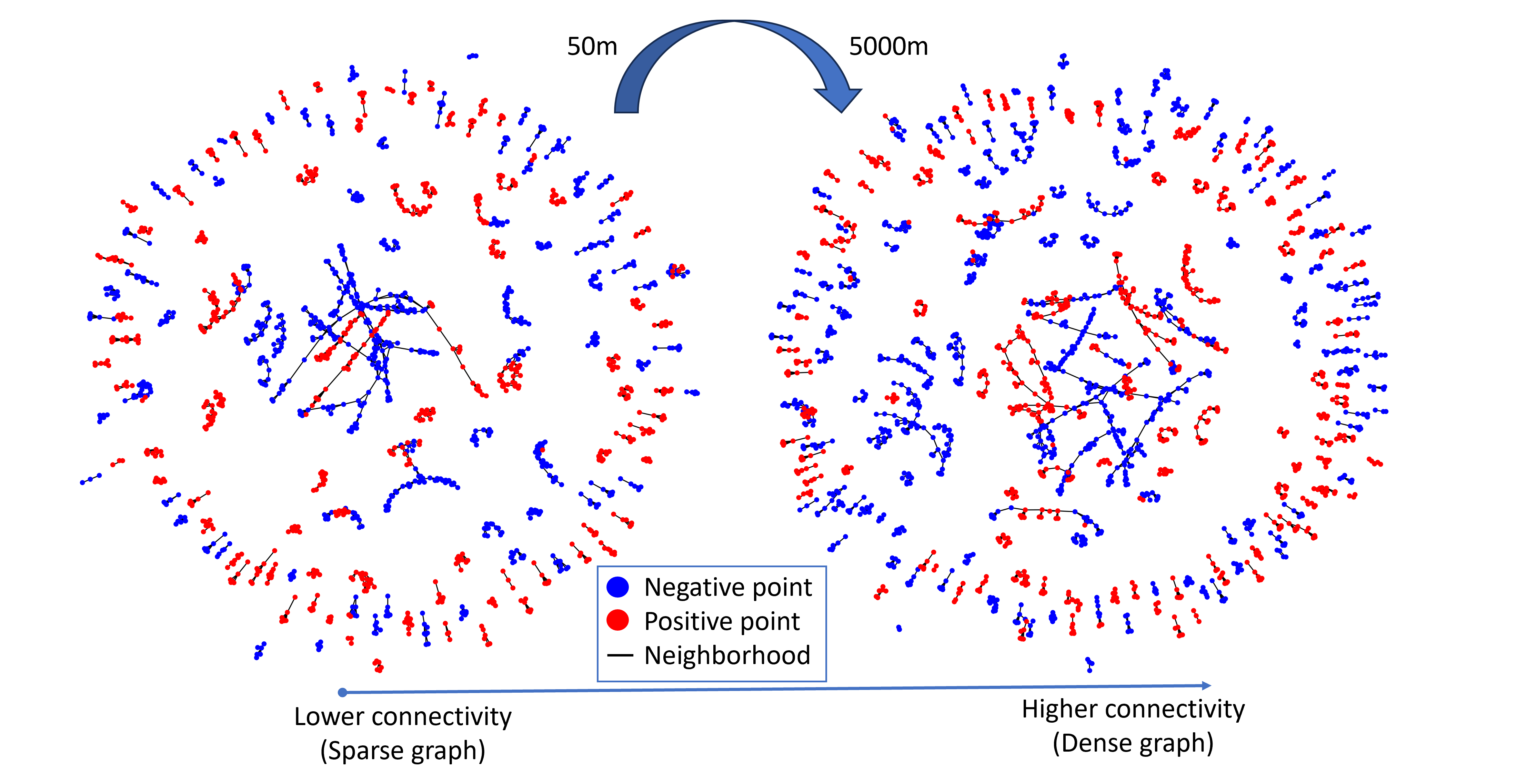}
  \caption{Graph of samples in the hard negative sample dataset and the connection with neighbors. Each node in the graph represents a data point; the node's color corresponds to its class; red indicates landmine presence (positive data point), and blue means clear of landmine (negative data point).}
  \label{fig:nx}
\end{minipage}
\end{figure}

We experiment three Neural Networks on the 18 features set dataset. In particular, we test Feedforward Neural Network (FNN), Graph Neural Network (GNN), and GNN adding weights (distance to the neighboring points). 
FNN treats each point as an independent individual, while GNN's two graph convolutional layers aggregate the features of the neighbors. 
Fig.~\ref{fig:mlmodel_comparison} compares the results of the tree neural network models with Random Forest. We can observe that GNN (simply connecting the neighbors) performs better than FNN in 500m and 5000m hard negative samples. Merely considering neighbors' connection (without the distance to them) does not help in 50m hard negative samples. This can be seen by comparing the plotted graphs in Fig.~\ref{fig:nx}\footnote{Fig. \ref{fig:nx} only shows a set of 3,000 data points in the graph for simpler visualization}. More nodes (data points) of different classes are connected in 5000m hard negative samples  because larger hard samples have a higher chance of linking to other landmine contamination areas. 

Another significant result from Fig.~\ref{fig:mlmodel_comparison} is that GNNs with weights outperform the other models for each dataset, especially when the sample is close (50m or lower). This implies that taking the distance to the five nearest neighbors and their features into account can help to predict the landmine presence probability of the point in question. 

\subsection{Error Analysis of the Study Area} 
\label{sec:error_analysis}
\begin{figure}[h]
\centering
  \includegraphics[width=\linewidth]{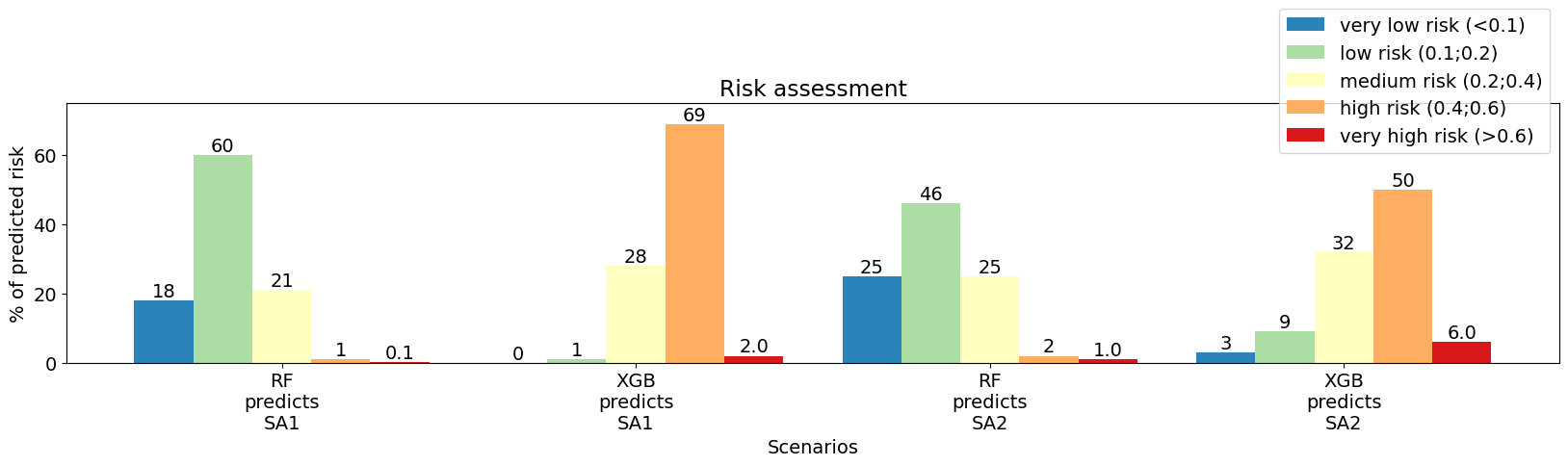}
  \caption{Percentage of five predicted risk assessment categories in the two study areas. The colors correspond to the landmine risk visualization on maps in Fig.~\ref{fig:SA1_RF_heatmap}-~\ref{fig:SA2_Xgb_heatmap_B}.}
  \label{fig:risk_assessment}
\end{figure}

\begin{figure}[h!]
\begin{minipage}{.23\textwidth}
  \centering
  \includegraphics[height=2.8cm]{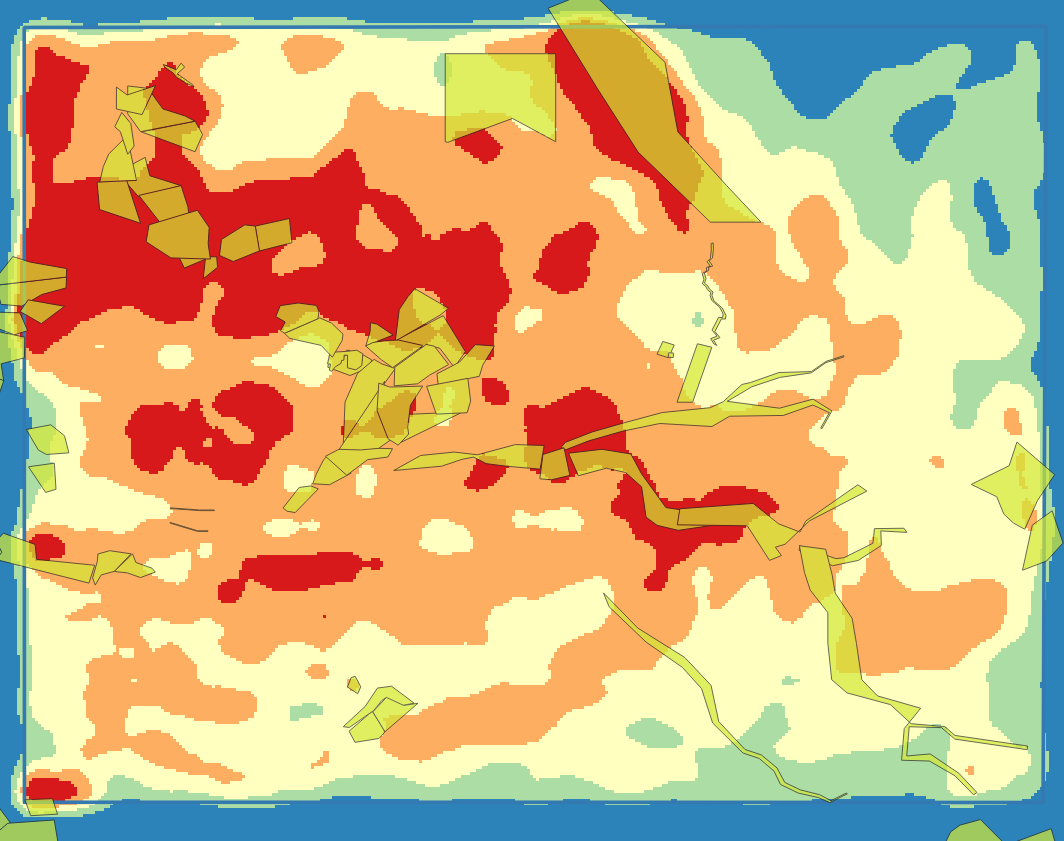}
  \caption{Prediction result of SA1 with Random Forest}
  \label{fig:SA1_RF_heatmap}
\end{minipage}
\hspace{0.5em}
\begin{minipage}{.23\textwidth}
  \centering
  \includegraphics[height=2.8cm]{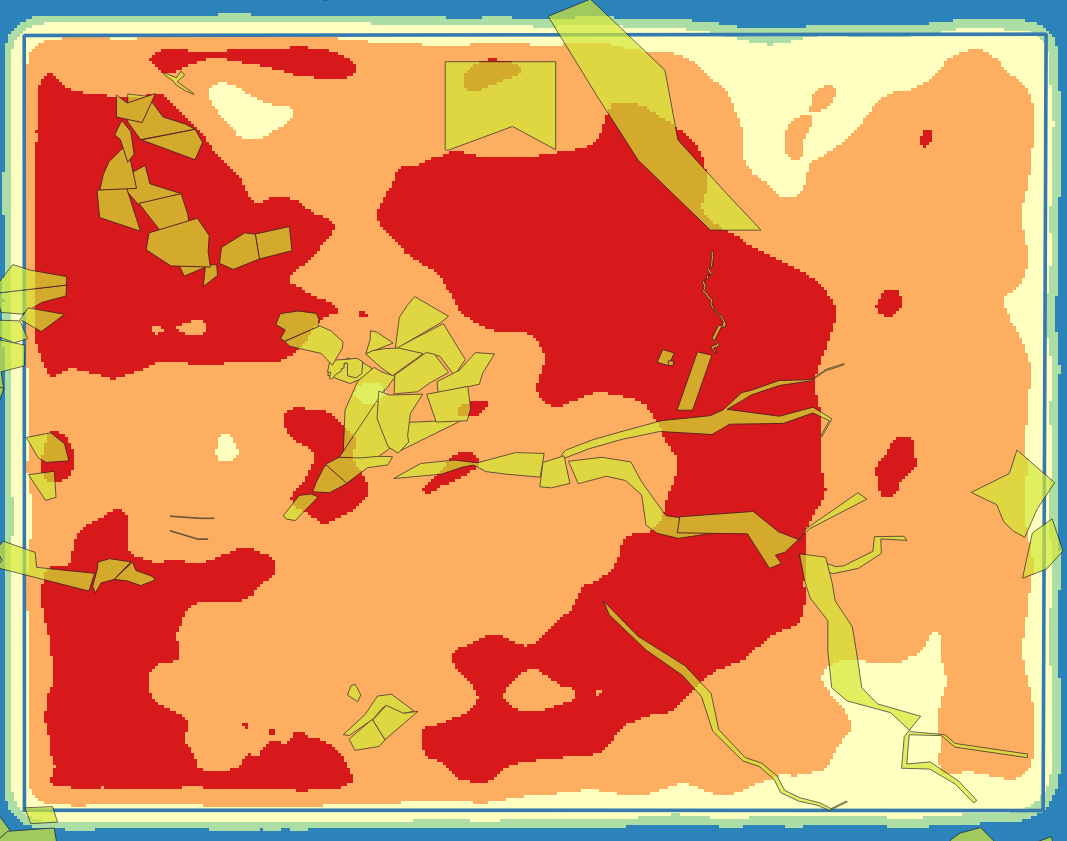}
  \caption{Prediction result of SA1 with XGB}
  \label{fig:SA1_Xgb_heatmap}
\end{minipage}%
\hspace{0.5em}
\begin{minipage}{.23\textwidth}
  \centering
  \includegraphics[height=2.8cm]{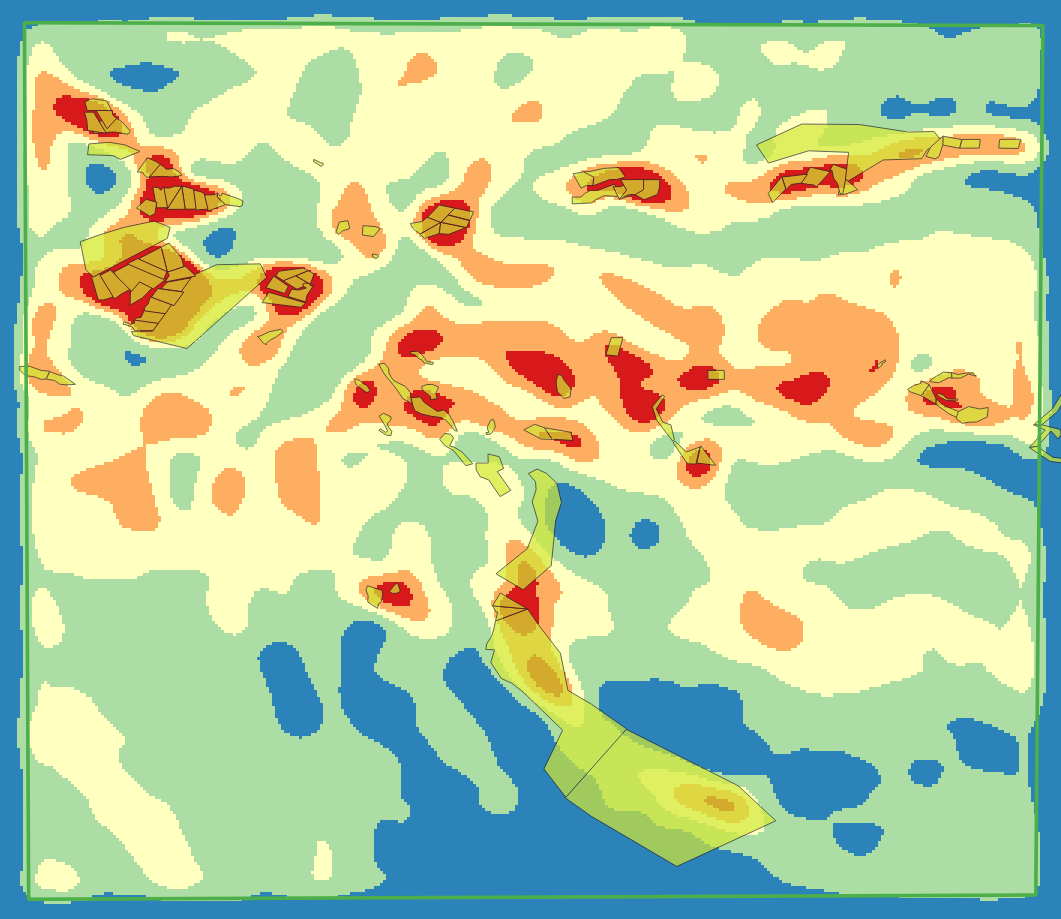}
  \caption{Prediction result of SA2 with Random Forest}
  \label{fig:SA2_RF_heatmap_B}
\end{minipage}
\hspace{0.5em}
\begin{minipage}{.23\textwidth}
  \centering
  \includegraphics[height=2.8cm]{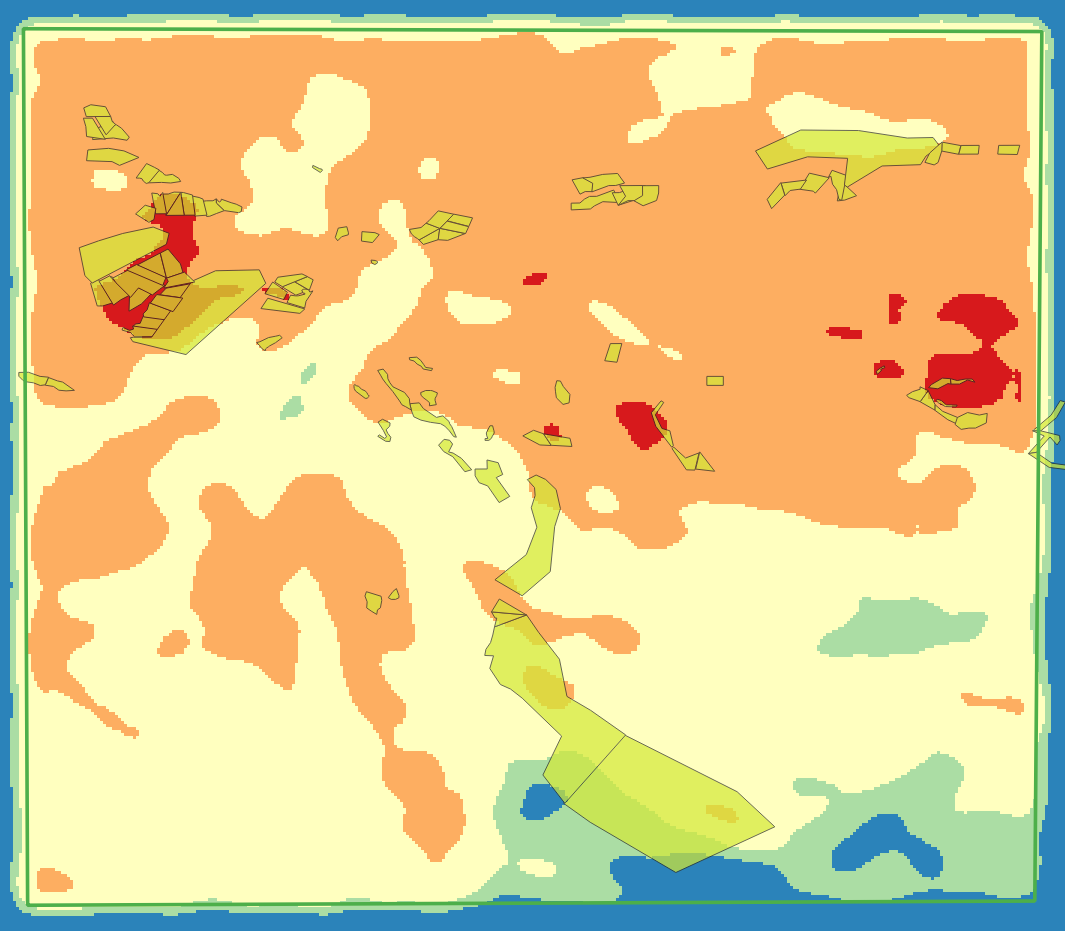}
  \caption{Prediction result of SA2 with XGB Forest}
  \label{fig:SA2_Xgb_heatmap_B}
\end{minipage}%
\end{figure}

In this section, we further investigate the models' prediction ability by plotting the landmine risk map in both study areas. Desk-AId is meant to augment already existing landmine desk assessment tools without disrupting already established workflow in the field. Thus, we use QGIS, an open source GIS tool already adopted by NGOs, for visualizing the results. Using the QGIS platform, predicted probability can be compared with the actual landmine distribution. We generated the scaled heat maps in Figures \ref{fig:SA1_Xgb_heatmap}-\ref{fig:SA2_RF_heatmap_B}, where the yellow polygons are the landmine contamination. We choose RF (weighted) and XGB trained on mix samples because their AUC scores are above 0.5 in both study areas, meaning they have discriminative prediction ability. Noted that in order to facilitate interpretation, the landmine risk maps have been classified into five risk levels and colored variously in the heatmap: very low (blue), low (green), medium (yellow), high (orange), and very high (red), based on the following cut-off values of probability: 0.1, 0.2, 0.4, and 0.6 (see Fig.~\ref{fig:risk_assessment}). The thresholds are set empirically. The same approach is also used in the work~\cite{Rafique2019} and \cite{Schultz2016}, where they specify that the threshold has to be evaluated and set by the mine action experts for an operational system. The percentage of points predicted probability in two study areas and the intervals' corresponding color in the risk map is summarized in Fig.~\ref{fig:risk_assessment}.

 
Comparing the risk map of SA1 in Fig. \ref{fig:SA1_Xgb_heatmap}-\ref{fig:SA2_RF_heatmap_B} and the statistics in Fig. \ref{fig:risk_assessment} provide a view that the discrimination achieved by the RF model is more effective within the study area. The XGB tends to have more false positives. From the summarized table, XGB predicts 71\% of the observations as high or very high risk, with landmine presence possibility higher than 0.4 (orange color in the risk map). This leads to almost the entire region having high landmine risk and could result in a rapid depletion of available mine action resources. For RF, the part where it is detected as high risk is mainly correlated with the landmine contamination area. Nevertheless, both models do not perform well in the southeast region. 

Figures~\ref{fig:SA2_Xgb_heatmap_B} and \ref{fig:SA2_RF_heatmap_B} compare the performance of the two models in SA2. As discussed previously, RF performs better in this region. It can detect the contamination regions as high risk (orange color on the map). XGB, on the other hand, still suffers from the problem of a high portion of false positives, and it does not detect the contamination area in the southeast region. 

The result from the study areas implies that in general RF is suitable for building a model that generalizes from the large region, such as the country's land, and subsequently predicting outcomes for specific study areas. Furthermore, RF is capable of delivering superior performance when the variability of features is extensive. On the other hand, LR and XGB could be helpful when the use case is, for example, validated inside the study region. In other words, if the demining operation is partly finished in the study region, LR and XGB can validate partially inside the area to avoid overfitting rather than cross-validation in the whole country land. This leaves an opportunity for future investigation.

\section{Discussion}


\textbf{Application for Demining Use Cases:} The Desk-AId system that we developed has proven to be responding satisfactorily to different scenarios. The final target is to have a system easily replicable in different regions/countries with minimal effort. 
For this purpose, we focus on the usage of features calculated on geographic and socio-economic information that are easy to find such as building locations, waterways, national/regional borders, and roads. The features used for the training are virtually available on every geographic location. Further, we also overcome the issue of lack of ground truth for negative data points that is a common problem for this kind of datasets. We experience that mixing the hard negative sampling approach for areas close by known areas and filling the rest of the country with the random sampling approach produce the best results. In addition, we see that, even if in some cases the F1 scores are not high such as the case of testing a simple classifier trained on small features set into uncharted areas (see Fig.~\ref{fig:random_vs_hardnegatives_sa1}-~\ref{fig:random_vs_hardnegatives_sa2}), the visualization of the risk prediction as continuous numbers matches quite well with the actual separation of hazardous areas with non-hazardous areas (see Fig.~\ref{fig:SA1_RF_heatmap}-~\ref{fig:SA2_RF_heatmap_B}).

\textbf{Field Trials:} Easy applicability of the Desk-Aid helped us conduct a successful proof-of-concept project with the open datasets from Afghanistan and received various positive feedbacks from the \textit{anonymized} humanitarian domain expert and their international humanitarian organization, mainly due to the following factors: 
\begin{itemize}
\item The innovative use of AI in the very large scale and providing initial results without high costs 
\item The usefulness of Desk-AId results in the desk assessment phase by the domain experts who are experienced in field testing and demining 
\item High potential for improving efficiency through prioritization and helping the decision-making for landmine clearance 
\item Potential reduction in demining time and the substantial costs coming with it (estimated as high as tens of billions of dollars)
\item Complementing existing innovations such as using drones for pinpointing surface-level mines 
\item Having a structured method and pipeline that are agnostic to the specific regions/areas or datasets
\item Building the technology on top of existing open-source components and standards that are accepted and used by the humanitarian community
\end{itemize}

In addition, we are already testing this Desk-AId pipeline in the Cambodia region and working for a field trial deployment in collaboration with the Cambodian Mine Action Centre (CMAC). Further, we are discussing with several NGOs and mine agencies for additional trials in few other post-conflict countries.

\textbf{Press Releases and Follow-up Research from the Community:} As a result of the proof-of-concept study, there have been several press releases in 2023 explaining the approach and developed technology behind the Desk-AId\footnote{Press Release-a, \url{http://www.nec.com/en/press/202303/global_20230329_01.html}}\footnote{Press Release-b, \url{https://www.nec.com/en/global/corporateblog/202309/01.html}}\footnote{Press Release-c, \url{https://www.nec.com/en/global/sdgs/innovators/project/article13.html}}. The released technology received global news coverage and raised attention in the humanitarian mine action community. As a result, several groups started applying the Desk-AId approach and started follow-up research activities~\cite{rubio2023reland}. The follow-up studies from relevant academia and organizations such as UNMAS~\cite{rubio2023reland} contribute to the non-technical survey problem and towards the solution to the global problem of demining.

\textbf{Technical Aspects of Desk-AId:}The granularity of the areas in the proposed system is a minimum of 50 meters. Prediction with a minimum 50 meters of granularity are reasonably unaffected by dynamic changes of the soil, also in regions with very high erosion after decades to the time of landmine displacement (as is the case of the coastline of the Skallingen peninsula~\cite{jebens2013analyzing} when mines placed during the second world war are cleared in 2012). Once the desk assessment is performed by our Desk-AId system, other techniques for non-technical and technical surveys (see Fig.~\ref{fig:demining_operation} might involve different instruments to collect new dedicated measurements such as airborne sensor for remote sensing~\cite{Jebens2021}.

We also explore the utilization of additional set of features, even if those might lead to higher computational complexity. In particular, we experience that for the simpler classifier models, this has alternate effects on different buffer zone sizes. However, the additional features have a positive effect for the application of a trained classification model to an uncharted area. The results are more promising and robust when we train and test with the graph neural network (such as GCNN) that takes into consideration the relation of points with each other (in our case 5 nearest neighbour points). The best performances are obtained when introducing the weights on the links between the graph nodes based on the distance between the points.

\textbf{Additional Techniques:} Further, we considered the application of different techniques for geospatial AI such as compressive sensing~\cite{li2020overview} that is meant to reconstruct a signal sampled at much lower frequency than the traditional Nyquist-Shannon sampling theorem. An issue is that such techniques are basically a convex optimization problem~\cite{li2020overview} that might require vast computational resources, thus, they are not best suited for democratization of AI~\cite{seger2023democratising}. Compressive sensing  Some work in state-of-the-art try to address this issue for spatial analytics~\cite{ayanzadeh2019compressive}. Another possible technique to explore is sparse coding~\cite{lavreniuk2022reviewing}. Also this technique is quite expensive in terms of computational costs since it translates again into a convex optimization problem.

We experience that each data feature we select do not have strong correlation with the presence of an hazard in the area. This is shown in the feature correlation matrix of Fig.~\ref{fig:correlation_matrix}. However, we implemented models that are capable of calculate a risk assessment and they are useful for the domain experts (see Fig.~\ref{fig:SA1_RF_heatmap}-~\ref{fig:SA2_RF_heatmap_B}). A possible future work is to include explanability measures to the risk assessment prediction to provide as much information as possible to the domain experts during the desk assessment phase. 

\textbf{Ethical Use and Risks:} The problem addressed in this article requires also a Dual-Use Research of Concern (DURC) discussion. It is undoubtful that high-accuracy predictions in automatic landmine risk assessment can greatly improve the outcome of desk assessment phase. As a result, a more educated plan and prioritization for fast clearance of landmines fosters human life safety and socio-economic growth. Nevertheless, such systems might be used in a malicious way for different purposes such as military operations. Currently there is no procedure in place to avoid vicious usage of this type of technology. Our approach is to continuously discuss with several NGOs that actively work in demining operations and humanitarian action since decades. Further, we often participate to demining workshops (such as GICHD innovation conferences, workshops and webinars\footnote{https://www.gichd.org/the-gichd/events-training/}) to be transparent on the targeted application and receive possible feedback from landmine domain experts on ambiguous usages.

\section{Conclusion}
The deployment of landmines involves considerable aspects and typically stands for complex reasons or contingency behind. The process of demining operations starts from desk risk assessment of potential hazardous areas. Reaching good performance at this stage is crucial to efficiently spend time and resources at later stages.

In this work, we develop the Desk-AId as a tool for automatically detecting landmine risk in various different regions by exploiting the contamination across the whole country land and considering the features in the geographical and socio-economical domains, and historical reports. We explore different approaches to deal with the problem of covering vast areas (country-wide in our case) and the application of a trained ML model to an unseen area. We propose a hard negative data sampling strategy that selects the informative negative points and compare it with randomly sample strategy where points are selected on the whole available land. 
Moreover, we not only analyze the correlations and multicollinearity between those features but also their roles present in predicting landmine contamination. 
To this extent, two tasks of HMA are conducted from the generic models: the first one distinguishes the vicinity of the contamination area by comparing the well-established to the state-of-the-art models, which utilize location-based graph structures defined in this work to build the GCNNs models. It has shown an outperforming result in aggregating the neighboring information. The second task successfully detects the landmine risk in previously unseen regions, for which a heat map is generated from the predicted probability. The size of hazardous area is significantly reduced from the RF model prediction and is therefore highly practical in the HMA usage for non-technical mine action experts. Besides qualitative assessment, each of the experiments is evaluated quantitatively with two sets of attributes and distinct negative sample strategies. 

For the future work, we plan to explore a wider range of data collection from open sources, and apply the proposed pipeline in the other countries, using the proposed pipeline of Desk-AId to help plan the humanitarian operations and help solving the global demining problem.

\section*{Data availability statement (DAS)}


The data that support the findings of this study are available in [repository name] at [URL/DOI], reference number [reference number]. These data were derived from the following resources available in the public domain: [list resources and URLs]

\bibliographystyle{tfv}
\bibliography{minedetection-ref}

\end{document}